# Infrared Properties of Star Forming Dwarf Galaxies: I. Dwarf Irregular Galaxies in the Local Volume [1]

Vaduvescu Ovidiu, McCall Marshall L.

*York University, Department of Physics and Astronomy*
*4700 Keele Street, M3J 1P3, Toronto, ON, Canada*
*email: ovidiuv@yorku.ca, mccall@yorku.ca*

Richer Michael G.

*Observatorio Astronomico Nacional, Instituto de Astronomia, UNAM,*
*PO Box 439027, San Diego, CA 92143-9027, USA*
*email: richer@astrosen.unam.mx*

Fingerhut Robin L.

*York University, Department of Physics and Astronomy*
*4700 Keele Street, M3J 1P3, Toronto, ON, Canada*
*email: rfinger@yorku.ca*

## ABSTRACT

A sample of 34 dwarf irregular galaxies in the Local Volume, most nearer than 5 Mpc, has been imaged in the near-infrared (NIR) in $J$ and $K_s$ at CFHT in Hawaii and OAN-SPM in Mexico. Absolute magnitudes in $K_s$ range from -14 to -18. In CFHT images, stars brighter than $M_{K_s} \sim -7.5$ were resolved. We show that the resolved component comprises more than 50% of the light from star formation bursts within the last 3 Gyr. In most cases, the resolved population down to $M_{K_s} = -7.5$ represents less than 5% of the total NIR flux in $K_s$, with fractions in $J$ being $1.5 - 2$ times larger. Thus, the NIR light of dIs can be considered to be predominantly contributed by stars older than about 4 Gyr. Although exponential at large radii, surface brightness profiles for the unresolved component flatten in the centres. They can be fitted across the whole range of radii with a hyperbolic secant (sech) defined as a function of two parameters: the central surface brightness and the scale length of the exponential. With respect to this model, only two galaxies (NGC 1569 and NGC 3738) show an excess of flux in the centre, both of which are hosting starbursts. Isophotal, total, and fitted sech magnitudes have been calculated for all galaxies for which the unresolved component was detected, along with semimajor axes at $\mu_J = 23$ mag arcsec$^{-2}$ and $\mu_{K_s} = 22$ mag arcsec$^{-2}$. The scale length and the semimajor axes correlate linearly with absolute isophotal magnitude. The same is true for colors and the central brightness. More luminous dIs tend to be larger, redder and brighter in the centre. The fraction of light contributed by young stars is independent of both luminosity and central surface brightness. The Tully-Fisher relation shows considerable scatter, but residuals are tied to surface brightness. The galaxies appear to lie in a "fundamental plane" defined by the sech absolute magnitude, the sech central surface brightness, and the HI line-width. The rms of residuals in $M_K$ is only 0.4 mag, which implies that the plane can be used to evaluate the distances of star-forming dwarfs. Corrections for tilt do not reduce the residuals, so line-widths must be governed predominantly by random motions. Color-magnitude diagrams are presented for 29 galaxies in which stars were resolved. Most show a finger centered around $J-K_s = 1$ mag. In some cases, there is a red tail extending to $J-K_s = 2.5$ mag. Most color profiles constructed for the unresolved component show a remarkably constant $J - K_s = 0.8$ to 1.0 mag, matching the color of the finger in the CMDs.

*Subject headings:* galaxies: dwarf irregulars, surface photometry, fundamental parameters; infrared: galaxies, stars; color-magnitude diagrams;

## 1. Introduction

Dwarfs are the most numerous galaxies in the Universe. Defined as having absolute magnitudes fainter than $M_V \sim -18$ mag (Grebel 2001), dwarf galaxies are organized in four groups based on their optical appearance: dwarf ellipticals (dEs), dwarf spheroidals (dSphs), dwarf irregulars (dIs), and blue compact dwarfs (BCDs).

Among dwarfs, dIs and BCDs are considered the building blocks for more massive galaxies in hierarchical merger scenarios. As such, they are important probes for studying matter in its near-primordial state. Despite much work in the last two decades, many questions remain open about dwarf galaxies, the most important being the evolutionary links between dIs, BCDs, and dEs. Some authors suggest that BCDs and dIs are the same type of galaxy, with BCDs being dIs undergoing bursts of star formation (e.g. Thuan 1985). Others argue that dIs are a fundamentally distinct type from BCDs, with no simple evolutionary links between them (e.g. James 1994; Richer & McCall 1995). Papaderos et al. (1996) review previous arguments on the evolutionary relationships between BCDs, dIs and dEs without reaching firm conclusions.

---

[1]Based on observations obtained at the Canada-France-Hawaii Telescope (CFHT) which is operated by the National Research Council of Canada, the Centre National de la Recherche Scientifique de France, and the University of Hawaii, also based on data acquired at OAN-SPM Mexico.

Having a low surface brightness ($\mu_B \geq 22$ mag arcsec$^{-2}$), dIs are gas-rich galaxies showing no nucleation and an irregular optical appearance usually dominated by scattered H II regions associated with star formation. The proportion of metals in dIs is very low, with metallicities (Z) ranging between $Z_\odot/40 - Z_\odot/3$ (Kunth and Ostlin 2000). Skillman, Kennicutt & Hodge (1989) confirmed a strong correlation between absolute magnitude and metallicity for local dIs, in the sense that the least luminous are the most metal poor. The existence of a relation between metallicity and luminosity constitutes a manifestation of something deeper which needs to be understood in order to address the chemical evolution of dwarf galaxies (Kunth and Ostlin 2000).

According to Kunth and Ostlin (2000), the origin of the concepts of "compact galaxies" and "blue compact galaxies" (BCGs) is due to Zwicky (1965). Those BCGs which are less luminous than $M_B \approx -17$ are commonly referred to as blue compact dwarfs (BCDs). BCDs include gas, stars, and usually starburst regions, all of which are centrally concentrated. Due to their pronounced compact starbursts, BCDs have higher surface brightnesses ($\mu_B \leq 22$ mag arcsec$^{-2}$). Like dIs, BCDs are also metal poor, with $Z$ as low as $Z_\odot/50$, showing a trend of redder colors toward higher metallicities (Thuan 1983).

Most of our knowledge about dwarfs comes from members of the Local Volume, i.e., our neighbourhood within about 10 Mpc (Karachentsev et al. 2004), whose stellar content can now be studied in detail using large ground-based facilities or the HST. Additional information comes from studies of the global properties of dwarfs in clusters.

The utility of light as a gauge of galaxy mass depends upon the wavelength of observation. In star-forming galaxies like dIs and BCDs, the young population shines brightly in the visible, overwhelming the light from the old stellar component, which constitutes the bulk of the mass. This problem can be minimized in the near-infrared (NIR), where the intermediate-age and old populations become more visible. Furthermore, light in the NIR is much less attenuated by extinction than light in the visible. More data in the NIR has become available with the advancement of NIR technology in the last decade. NIR photometry is now available from deep studies of selected objects as well as from shallower but more homogeneous surveys like 2MASS or DENIS. Despite these efforts, well-defined stellar populations like the asymptotic giant branch (AGB) and the red giant branch (RGB) have been identified in the NIR in only a few close dIs, such as IC 10 (Borissova et al. 2000a), IC 1613 (Borissova et al. 2000b), NGC 1569 (Aloisi et al. 2001), NGC 4214 (Drozdovsky et al. 2002), NGC 1705 (Tosi et al. 2001), the Magellanic Clouds (Nikolaev & Weinberg 2000; Cioni & Habing 2003), and NGC 6822 (Cioni & Habing 2004). Thus, more work needs to be done in this direction.

In §2, we present the selection criteria for our observing sample. In §3, we describe our observations. The image reduction method is presented in §4, followed by the photometric calibration in §5. In §6, the stellar photometry is addressed and the catalogs of selected stars from which color-magnitude diagrams were constructed are described. Surface photometry is presented and analyzed in §7. In §8, NIR properties of the dIs are discussed, based on correlations between their sizes, central surface brightnesses, unresolved fluxes, and the ratios of the flux from the resolved light to the total light. Distances taken from the literature are homogenized to a common distance scale, used throughout the paper, and presented in the Appendix. Conclusions are presented in §9.

## 2. The Sample

In previous studies of dwarf galaxies in the field (Lee, et al. 2003; Richer & McCall 1995), samples were defined based on two criteria:

1. detection of the [O III]$\lambda 4363$ emission line for the determination of oxygen abundances;

2. a reliable distance (i.e. determined from stellar constituents, such as Cepheids and the TRGB).

   The observing facilities to which we had access imposed two additional criteria on our sample:

3. an apparent size smaller than $\sim 3'$, possible to image by both facilities used for this work;

4. target visibility from each facility during the observing time granted.

We identified 34 dwarf galaxies which satisfy these conditions. They are listed in Table 2.

## 3. Observations

Between 2001-2004, we had 6 observing runs (27 nights in total) during which we observed 34 field dwarfs in the NIR. Of these, a total of about 12 nights were used to observe the field dIs. The remaining 15 nights were dedicated to BCDs in the Virgo Cluster. All allocated observing time was bright (Moon within 5 days of full). Our observations of the field dIs are presented in this paper. The observations of the Virgo BCDs will be included in a future paper.

We obtained our observations using the 3.6 m Canada-France-Hawaii Telescope (CFHT) located atop Mauna Kea, Hawaii, and the 2.1 m telescope of the "Observatorio Astronómico Nacional in the Sierra San Pedro Mártir" (OAN-SPM) in Baja California, Mexico. Due to cloudiness (especially in Mexico), the observing efficiency was about half of the allocated time. Consequently, data for field dIs were acquired on about 6 nights in total (3 at CFHT, and 3 at OAN-SPM). A log of our observations of the 34 field dIs is presented in Table 2.

Observing faint extended sources in the NIR is difficult because the sky is very bright in the NIR, with an average surface brightness of $\mu_K \sim 13\,\mathrm{mag\,arcsec}^{-2}$ for a dark site like Mauna Kea[2]. Furthermore, the surface brightness of the isophote defining $D_{25}$ in the $B$ band corresponds to 21 to $23\,\mathrm{mag\,arcsec}^{-2}$ in $K$ (e.g., Fioc & Rocca-Volmerange 1999). In the $K$ band, this faint surface brightness makes galaxies 8 to 10 mag fainter than the sky! An additional problem is the background due to the atmosphere, detector, and thermal radiation, which vary both temporally and spatially. Thus, conservative approaches to observing and reductions were taken to ensure accurate background subtraction, the details of which are given below (see also Vaduvescu & McCall 2004).

### 3.1. CFHT Observations

On three nights between Feb 28 − Mar 3, 2002 we observed 26 field dIs in $J$ and $K_s$ at the 3.6 m $f/8$ Cassegrain focus of CFHT. We employed the CFHT-IR camera which is equipped with a HgCdTe array with $1024 \times 1024$ pixels. The scale was $0\farcs211\,\mathrm{pix}^{-1}$, yielding a $3\farcm6 \times 3\farcm6$ FOV.

Two other nights between Mar 6 − 8, 2004 were used to observe three more dIs as part of another project (Fingerhut et al., 2005 in preparation). For these 3 targets, we employed

---

[2]Cf. Vanzi & Hainaut (2002), online at European Southern Observatory, http://www.eso.org/gen-fac/pubs/astclim/lasilla/l-vanzi-poster

the same instrument described above.

In 2002, conditions were photometric on all three nights, with an average seeing of $0\rlap{.}''5$ in $K_s$, humidity less than 30%, and a steady temperature of about $-2°C^3$, all typical of Mauna Kea when weather is good. The two nights in 2004 were photometric, with the temperature dropping from $+2°$ C to $-4°$ C and the humidity oscillating between $10-30\%$ for most of the run.

One of the main objectives of the present project is to disentangle the young stellar population from the old. We aimed to resolve young stars down to an absolute magnitude $\mu_{K_s} = -7.5$ mag, i.e. AGB stars younger than roughly 8.5 Gyr (Bressan et al. 1994) (see their Fig. 25 and Table 2), assuming a metallicity of $Z = 0.001$. Calculated at a distance of 5 Mpc (the limit for most of our targets), this absolute magnitude limit corresponds to an apparent magnitude $K_s = 21$ mag. We also aimed to detect isophotes as faint as $\mu_{K_s} = 23$ mag arcsec$^{-2}$ and $\mu_J = 24$ mag arcsec$^{-2}$, i.e., as faint out as $\mu_B = 25$ mag arcsec$^{-2}$. In order to reach these limits, we used total exposure times of 10 min in $K_s$ and 5 min in $J$. In 2002, this was obtained by combining individual exposures of 60 sec in $K_s$ and 100 sec in $J$. In 2004, we combined exposures of 75 sec in $K_s$ and 60 sec in $J$.

Based on a statistical analysis that we performed on a representative data set (800 images taken in $K_s$ and $J$ during three photometric nights using CFHT-IR camera in 2002), we derived some results regarding the variation of the sky background (Vaduvescu & McCall 2004). At CFHT, the *background level* (defined as the median signal in any given sky frame) varies from frame to frame by about 0.5% per minute in $K_s$ and 0.7% per minute in $J$, on average. The *background pattern* (defined across the $3\rlap{.}'6 \times 3\rlap{.}'6$ FOV by the range of count levels in frame differences) varies on average by about 0.2% per minute in $K_s$ and 0.3% per minute in $J$.

In order to detect the outermost regions of the dwarfs, which is where the old population was expected to make the largest contribution to the light, we sampled the sky as often as the target, using the same exposure time for both target and sky. This was done in order to properly account for temporal variations in the NIR background. The spatial signature was removed using a flat field derived for each filter from twilight sky images taken each night or morning.

We employed the following observing sequence, sampling the sky in a region 4–5′ away from the target:

$$sky - target - sky - target - sky - ... - sky - target - sky \qquad (1)$$

---

[3] Cf. UKIRT Graphical Weather Server, online at http://www.ukirt.jach.hawaii.edu/weather_server/weather.cgi

Sky frames were observed in the same region of sky as each galaxy, namely 4–5′ North or South of the target. To improve the identification of bad pixels and to remove small contaminants, we adopted the classical approach of dithering each new sky frame and each new galaxy frame by 5–10″ with respect to the previous sky or galaxy frame.

### 3.2. OAN-SPM Observations

Between 2001-2004, we were granted four observing runs for a total of 21 nights at the $f/4.5$ focus of the 2.1 m telescope at OAN-SPM. In all runs, we used the CAMILA NIR camera, which is equipped with a NICMOS3 256 × 256 pixel array (Criz-Gonzales et al. 1994). The scale was $0\rlap{.}''85\,\text{pix}^{-1}$, yielding the same $3\rlap{.}'6 \times 3\rlap{.}'6$ FOV as the CFHT.

Unfortunately, the weather in Feb-Mar was unstable throughout the four runs at OAN-SPM (Tapia 1983). We had clear skies less than half the allocated time with only one night with photometric conditions. The average seeing was about $2\rlap{.}''5$, but this does not degrade the surface photometry of our targets.

The primary objective of the OAN-SPM runs was to obtain deep imaging of a sample of Virgo BCDs in the NIR, in order to compare their NIR properties with those of field dIs (to be addressed in a future paper). At OAN-SPM, we imaged in total 10 dIs from the Local Volume, during the first part of the available clear nights. Total exposure times of 20–40 min in $K_s$ and 15–20 min in $J$ were used at OAN-SPM in order to match the CFHT surface brightness limit, taking into account the ratio of the two apertures (about 3) and the differences in the sky transparency. To obtain these total exposures, we combined individual frames exposed 1 min each in both $K_s$ and $J$.

Due to the larger total exposure time required at OAN-SPM, we sampled the sky half as often as the target in a region 3–5′ away from the galaxy. For all our OAN-SPM runs we alternated each sky frame with two target frames, adopting the following observing sequence:

$$sky - target - target - sky - target - target - sky - ... - sky - target - target - sky \quad (2)$$

We dithered each galaxy frame by 3″. The sky frames for each galaxy were observed at positions around the galaxy, 3–5′ away.

## 4. Data Reductions

### 4.1. Overview

One bad pixel map was built for each of our runs using flat field images taken with two different exposure times. It was applied as the first step in the data reduction process to all CFHT and OAN-SPM raw images (using the BADPIX task of IRAF).

Removal of the background represents the most important step in NIR observations, especially when observing faint extended sources (Vaduvescu & McCall 2004). Without proper care, variations between the sky and the galaxy frames can end up altering the faint regions of a galaxy, affecting the surface brightness profile and total flux. A cleaned sky frame was built for each galaxy frame from neighbouring sky frames taken immediately before and after the galaxy frame. The sky frame was then subtracted from the galaxy frame. The reduced galaxy frames were corrected by the flat field, then leveled and median-combined using IMCOMBINE in order to eliminate hot pixels, cosmic rays, and any "black holes" remaining in the reduced frames due to imperfections in the cleaning of sky frames.

For the image reductions, we used a collection of IRAF scripts named REDNIR.CL. These were written by the first author and tailored to the different observing sequences used at CFHT and OAN-SPM[4].

### 4.2. CFHT Reductions

To generate the flat field, tracked but dithered exposures of the twilight sky were acquired each night and morning. For each filter, a sequence of equal-duration exposures was taken as the twilight brightened or faded. How the signal in any given pixel correlated with the mean for all pixels was used to quantify the relative sensitivity of that pixel to light, and thereby generate a flat field frame.

Often, small extended sources in the background were embedded in raw sky images, $8 - 10$ mag lower than the sky noise. The sky images were also contaminated by bright foreground stars. Vaduvescu & McCall (2004) developed a method to remove these contaminants. Using this algorithm, each pair of raw sky frames straddling a given galaxy frame was used to produce a "cleaned sky frame". For each raw sky frame, its dithered counterpart was subtracted in order to remove noise and reveal the stars and the extended sources. These sources were detected using DAOPHOT (Stetson 1987, 1992), then cleaned with IMEDIT. Finally, the subtracted sky frame was added back, in order to fill in the "black holes" created by stars in the subtraction, generating a cleaned sky frame. The final sky frame was built by averaging the two cleaned sky frames, in order to interpolate in time the background at the time of observation of the galaxy.

---

[4]The REDNIR.CL NIR image reduction scripts available at http://www.geocities.com/ovidiuv/astrsoft.html

Fig. 1 presents the reduced $K_s$ images of the 22 field dIs observed at CFHT that show an unresolved component. The mosaic in Fig. 2 shows the 6 galaxies that do not show an unresolved component. For these six galaxies, we were unable to fit light profiles and derive total magnitudes. An additional target (Leo A) was not included in the figure, due to a bright nearby star whose spikes affected the reduced image. This galaxy did not show an unresolved component across the field. In order to emphasize subtle surface details for display purposes, we binned all frames by $4 \times 4$ pixels ($0.84 \times 0.84''$) using the BOXCAR function in IRAF.

### 4.3. OAN-SPM Reductions

Due to the smaller aperture of the OAN-SPM telescope, the lower sensitivity of CAMILA, and the higher number of frames per target (15–40 at OAN-SPM versus 3–10 at CFHT), we employed a simplified sky subtraction technique to reduce our OAN-SPM images. We simply subtracted the neighbouring sky frame from each galaxy frame, after leveling the signal in the corners to match the signal in the corners of the galaxy frame. Thus, the reduced galaxy frames are leveled using the corners, and median-combined into the final galaxy frame.

For each filter, a flat field was built from a high-signal and a low-signal flat from two series of twilight sky images taken at different illumination levels. The two flats were then subtracted in order to produce the final flat field image.

In Fig. 3, we present the reduced $K_s$ images of the 9 field dwarf galaxies observed at OAN-SPM.

### 5. Calibration of Photometry

### 5.1. CFHT Photometry Calibration

In 2002 at CFHT, we observed 4 to 6 Persson standard stars (Persson, et al. 1998) at different air masses and times of the night. The sky was photometric on all nights, allowing a determination of the transformation of instrumental magnitudes using observations of these stars. We assumed the following transformation equations:

$$k_S = K_s + k_1 + k_2 X + k_3(J - K_s) \qquad (3)$$
$$j = J + j_1 + j_2 X + j_3(J - K_s) \qquad (4)$$

Here $K_s$ and $J$ stand for the apparent magnitudes in the two bands, $k_S$ and $j$ the instrumental magnitudes, $k_1$ and $j_1$ the nightly zero-points, $k_2$ and $j_2$ the extinction coefficients, and $k_3$ and $j_3$ the color coefficients. Using the PHOTCAL package of IRAF, we tried two different approaches to determine the calibration coefficients in Equations 3 and 4.

In the first approach, we derived common extinction and color coefficients for the whole run, while determining the individual zero-points for each night. The extinction coefficients were 0.065±0.011 and 0.055±0.013, and the color terms were 0.039±0.005 and 0.003±0.005, in $K_s$ and $J$, respectively. The three nightly zero-points agreed within 0.014 mag in $K_s$ and within 0.008 mag in $J$, while the standard errors in the zero-points were less than 0.02 mag for all nights in both bands.

To check the uncertainties, we calculated the zero-points for 14 of our science fields using common 2MASS stars appearing in the fields (4-18 stars in each frame). The 2MASS stars were selected using the following conditions:

1. Catalog photometric uncertainties $\lesssim$ 0.15 mag (mostly $\lesssim$ 0.05 mag);

2. Point-like appearance on the CFHT images;

3. Magnitudes $K \gtrsim 13$ and the residuals $\delta K \lesssim 0.1$ mag.

We found that our original zero-points (with only one exception in 28 cases) were higher than the 2MASS ones by an average of 0.31 mag in $K_s$ and 0.24 mag $J$, reaching a maximum difference of 0.5 mag! To understand the inconsistency, we attempted another approach to derive the calibration coefficients.

In the second approach, solutions for the zero-points were decoupled from those for extinction and colour coefficients. First, we started by assuming initial values for the extinction coefficients of 0.05 in $K_s$ and 0.11 mag in $J$ for all nights, which are standard figures expected for spring at Mauna Kea (e.g. Manduca & Bell 1979; Guarnieri, Dixon & Longmore 1991). Dropping the color terms, we calculated individual zero-points for each standard star and each observation, then their average for each night. Next, we used the estimates for the nightly zero-points to calculate average global values for extinction and color coefficients for all three nights. Finally, we used the new estimates of the color coefficients to improve the individual nightly zero-points. Note that a similar treatment is suggested as a photometric calibration strategy for 2MASS data.

Using the second approach, we found extinction coefficients of $0.048 \pm 0.024$ and $0.109 \pm 0.025$ in $K_s$ and $J$, respectively, which are very close to the initial values. The color terms were $0.023 \pm 0.037$ in $K_s$ and $0.022 \pm 0.038$ in $J$. The three zero-points varied by 0.132 mag in $K_s$ and by 0.162 mag in $J$. Compared to the 2MASS zero-points, our values were higher by an average of 0.04 mag in $K_s$ and 0.08 mag in $J$. However, the difference was as high as 0.25 mag in one case.

During the 2MASS[5] survey, one calibration field was observed approximately every hour or two each night. Each observation provided six independent measurements of the standard stars. In the first two years of the survey, each standard star was measured about 900 times over approximately 50 nights. Based on these data, a global photometric calibration solution was published, which was used to reduce the 2MASS catalog (Nikolaev et al. 2000). Color terms were dropped and global extinction coefficients were calculated by linking data from different nights on a monthly basis. The nightly zero-points were assumed to vary linearly with time – see the example in Fig. 1 of Nikolaev et al. (2000), in which variations up to about 0.1 mag in range over the night were measured during a given night!

Given the higher frequency with which 2MASS standards were observed, also the compensation for zero-point variability, we decided to adopt the 2MASS frame zero-points to reduce all of our CFHT science fields. In seven cases (i.e., DDO 187, Holmberg IV, NGC 4163, NGC 4190, NGC 4789A, GR 8, and Markarian 209), when our CFHT frames were sparsely populated by 2MASS stars (fewer than four 2MASS stars per field), we interpolated the zero-points using neighbouring frames with more stars which were observed within one hour of the target, taking care to correct for the difference in the airmass. Note that extinction corrections were small, though, due to both the low airmasses ($\lesssim 1.5$) at which we observed most of the time and the small atmospheric extinction coefficients.

Based on the scatter in zero-points derived from different stars in a given frame, and taking into account the comparison between our second approach and the 2MASS zero-points, we expect most of our zero-point errors to be less than 0.05 mag. At maximum, errors may reach 0.1 mag in the following sparsely populated fields, in which we used less than four 2MASS stars: Mark 178, NGC 4163, NGC 4190, Mark 209, NGC 4789A, GR 8, Ho IV, and DDO 187.

---

[5]Cutri et al., 2003, 2MASS Data Release Documentation, online at: http://www.ipac.caltech.edu/2mass/releases/docs.html

## 5.2. OAN-SPM Photometry Calibration

As only one night at OAN-SPM was photometric, the zero-points were determined for each frame using common stars observed by 2MASS. We found average residuals of less than 0.04 mag in $J$ and less than 0.07 in $K_s$ for the 3-12 2MASS stars in each field. Based on an error analysis like that described for CFHT, we conclude that most of our OAN-SPM fields have zero-point errors less than 0.05 mag, with the maximum reaching up to 0.1 mag in four sparsely populated cases: UGC 4998, UGC 5979, UGC 5848, and UGC 8508.

## 6. Stellar Photometry

As specified above, our CFHT run was tuned to detect AGB stars younger than $\sim 8$ Gyr, out to 5 Mpc. By resolving the stars to this age limit, we sought to quantify how much the total light represents the old population, which constitutes the bulk of the mass. Due to the much poorer spatial resolution at OAN-SPM, we did not attempt to do stellar photometry based on the OAN-SPM data.

The ratio of the resolved to the total NIR light of a galaxy is a gauge of the contamination of the total light by young stars. To separate the resolved and unresolved components, we used KILLALL, an IRAF script built around the DAOPHOT package, which was developed and used previously in the visible by Buta & McCall (1999). This script iteratively identifies, fits and removes point-sources superimposed on a variable background. We applied KILLALL in a three-step process to remove the resolved sources, adopting a threshold level of five sigma for the first two runs and three sigma for the third run. In addition, we employed two PSF images, the first obtained using an analytic PSF model and the second using a quadratic variable. Finally, a few residuals from bright stars, spikes, and foreground extended sources were cleaned out manually using IMEDIT.

KILLALL's final output consists of:

1. The unresolved galaxy image; and

2. The point-source catalog, which includes all the resolved sources in the frame (i.e., resolved stars in the galaxy, foreground stars, and nuclei of background galaxies.

Fig. 4 shows the reduced $K_s$ image of the dwarf irregular galaxy NGC 1569 along with the image of the unresolved component produced by KILLALL[6].

---

[6]A movie showing the full KILLALL process is available on-line here:

## 6.1. Analysis of Crowding and Completeness

In a few galaxies observed at CFHT (NGC 1569, NGC 3738, NGC 4163, NGC 4190 and NGC 5264), the resolved component is crowded, especially near the galaxy centers. Consequently, some stars may be blended and magnitudes may be uncertain. A traditional method to judge the effect of crowding on counts and magnitudes is an artificial star analysis. It was originally developed for the case of a star cluster (Stetson & Harris 1988) and implemented in IRAF under the DAOPHOT package (ADDSTAR task). One of the most exhaustive analyses was performed by Aparicio & Gallart (1995).

In our application of the method, artificial stars are injected into a reduced galaxy frame (denoted the *real frame*) at random positions. Their randomly-assigned magnitudes (named *injected magnitudes*) span a range as wide as the real stars measured in the real frame. The resulting frame (denoted *synthetic frame*) would be processed again using KILLALL with the parameters and the PSF image as was used originally. In the presence of crowding, the processing of the synthetic frame recovers only a fraction of the stars present in the real frame, plus some injected stars having measured magnitudes known as *recovered magnitudes*. To evaluate the crowding, the injected and the recovered magnitudes of the injected stars are cross-correlated in order to calculate the number of lost stars and the errors in photometry. The appropriate number of injected stars depends upon the stellar density of the area under investigation (i.e., the crowding), and should be a small fraction of the total number of stars detected in the real frame, to avoid self-crowding (e.g, 10%, cf. Aparicio & Gallart 1995). To improve the statistics, the procedure should be repeated until the total number of injected stars reaches 10 times the number of stars in the real frame.

We injected artificial stars into NGC 1569, the most crowded galaxy of those observed at CFHT. Over the whole frame, we detected 3300 stars in both filters, of which 2658 stars were associated with (within the body of) the galaxy. If the whole frame were filled with stars at a density equal to that within the area of the galaxy, there would be about 6000 stars in total. Thus, 600 artificial stars (10% of 6000) were injected at a time.

We performed the test for both $J$ and $K_s$. KILLALL was run once, using a threshold of five and a variable PSF, because most of the artificial stars could be found right from the first step. Most of the 600 stars injected in the synthetic frames were recovered in both filters: 494 stars in $J$ (82 %) and 523 stars in $K_s$ (87%). Fig. 5 shows the residuals between the recovered magnitudes and the injected magnitudes. Most of the residuals are less than 0.1 mag, the standard deviation being 0.06 mag in $K_s$ and 0.09 mag in $J$. Comparing these

---

http://www.geocities.com/ovidiuv_astro/NGC1569-killall.gif

numbers with the standard deviation of the photometry of stars in the real frame (0.14 mag in $J$ and 0.11 mag in $K_s$), we conclude that crowding has a negligible effect on our photometry.

We also evaluated the degree to which completeness due to crowding varies with radius. First, we selected an ellipse to define the galaxy border, inside of which 252 artificial stars were injected. Altogether, 207 were recovered in $J$ (82%) and 217 in $K_s$ (86%), i.e., almost identical percentages as found across the whole field. Second, we halved the ellipse semimajor axis and repeated the counting. Of 66 artificial stars injected, 46 stars were recovered in $J$ (70%) and 48 in $K_s$ (73%). We conclude that crowding does not play an important role in our photometry even in the most crowded of the fields in our sample.

### 6.2. Selected Star Catalog

Many of the stars in the point-source catalogue produced by KILLALL are not a part of the galaxy. With the aid of the PSELECT task of IRAF, a new catalog named the *selected star catalogue* was created. Three steps were necessary to build it.

1. In rare cases, bright nuclei associated with known stellar superclusters, such as in NGC 1569, were mistakenly identified by KILLALL as stars. A small ellipse centered on each knot was used to select these nuclei, excluding them from the selected star catalogue.

2. Stars superimposed upon the galaxy image were identified as those point-sources located inside the outermost ellipse fitted by STSDAS/ELLIPSE task run on the unresolved galaxy image. This step selects the resolved stellar population associated with the galaxy, plus a few foreground stars and artifacts caused by the noise.

3. Milky Way stars residing in the foreground of the galaxy were rejected either by visual inspection (applicable to targets observed at moderate and high latitudes for which very few foreground stars need to be eliminated) or by setting an average limiting magnitude. Bright sources showing a flatter light profile than the stellar PSF, which we associated with possible globular clusters or small background galaxies, were eliminated. Due to the small number of these sources, this step could be performed by visual inspection.

The total flux coming from the resolved part of the galaxy was computed by simply adding fluxes of all sources in the selected star catalog.

### 6.3. Color Magnitude Diagrams

For each galaxy, we employed the selected star catalog to build the color-magnitude diagram for the resolved sources. We used the TABLES/TMATCH function of IRAF to match stars. Sources in $J$ and $K_s$ lying within 1 pixel (0.211″, i.e., less than half of the average seeing) were paired based on their positions in the two band images and were considered to represent the same star.

Fig. 6 presents the color magnitude diagrams (CMDs) of all galaxies observed at CFHT. In each panel, a horizontal line marks the apparent magnitude corresponding to $K_s = -7.5$ mag, as calculated from the distance modulus listed in Table 1. Three of the 29 galaxies show CMDs with more than 1000 sources, while fifteen more have more than 100 stars. In the right corner of each CMD, we include two numbers. The first represents the number of stars from the selected star catalog (i.e., thought to be associated with the galaxy). The second number represents the percentage of Milky Way contaminants lying in the foreground of the galaxy. For example, for NGC 1569 there are 2658 resolved stars in the galaxy, of which about 10% are expected to be Milky Way contaminants.

Most of the stars seen in the CMDs have $J - K_s$ colors between 0.6–2.0 mag. The CMDs reveal two main features: a blue sequence similar to a vertical finger centered around $J - K_s = 1$ mag and a red tail going as red as $J - K_s = 2.5$ mag. The blue finger is best seen in the CMD of Cam A, DDO 47, Ho I, NGC 4163, NGC 4190, and NGC 5264. In the CMD of NGC 1569, the base of the blue finger is slightly fainter that an absolute magnitude $K_s = -7.5$ mag, but in Cam A it appears to extend about 2 mag below this luminosity. The red tail is most clearly discerned in the CMD of Sextans B. From the bulk of stars at $(J - K_s, K_s) = (1.0, 20)$, a sequence of stars arises, first tilted towards the upper right, to higher luminosities and redder colors at $(J - K_s, K_s) = (1.8, 17.5)$, then proceeds approximately horizontally to redder colors. Like the blue finger in Cam A, the red tail also begins about 2 mag below an absolute luminosity of $K_s = -7.5$ mag. Other CMDs with evident red tails are Cam A, NGC 4163, NGC 4190, and DDO 187. The first three of these demonstrate that these features are not mutually exclusive.

Compared with recent evolutionary models (Girardi, et al. 2000; Marigo, Girardi, & Chiosi 2003), the vertical finger agrees with the position of the oxygen-rich intermediate and old RGB stars with higher luminosities (plotted as blue dots in Marigo, Girardi, & Chiosi (2003)). The red tail, meanwhile, can be associated with the thermally pulsating assymptotic giant branch (TP-AGB). As will be found later, the unresolved component has colors very similar to the blue finger in these CMDs. (cf. § 7.6).

The errors in photometry can be evaluated by plotting the formal uncertainties of the magnitudes of stars in the selected star catalogs (listed by DAOPHOT) as a function of the apparent magnitudes. Fig. 8 shows the errors for NGC 1569, our most crowded target. Based on this diagram, the average error is about 0.2 mag, with a maximum of about 0.5 mag in both bands at the faint end ($m_J \sim 22$, $m_{K_s} \sim 21$). Similar distributions were observed for the other targets.

Obviously, the contamination of CMDs with Milky Way stars depends on the Galactic latitude $b$ of each field, included in Table 1. A first quantitative approach of the contamination can be done for the targets with latitudes greater than $+20°$. Assuming an average galaxy size corresponding to about 1/3 of the FOV area, the expected number of Galactic contaminants as faint as $K_s = 21$ (about $V = 23$) ranges from 3 to 23 stars (Bahcall and Soneira 1980). More contaminants are expected in the case of our seven galaxies having latitudes lower than $|20°|$.

A more qualitative approach to check the contamination can be done by considering the CMDs for the field stars selected to lay outside the galaxy, based on its size and shape (i.e., the ellipse centre, $r_{23}$, $e$, and $PA$). In Fig. 7 we present the CMDs for nine galaxies at galactic latitudes less than $|20°|$ or whose CMDs in Fig. 6 show a blue finger, in order to check this feature. The first figure in the right corner of the CMDs in Fig. 7 represents the number of stars in the frame which are not part of the selected star catalog (i.e., probably contaminants from the Milky Way). The second number represents the percentage of Milky Way contaminants lying in the foreground of the galaxy (i.e., the same second number as in Fig. 6). The fractions were calculated based on the galaxy area relative to the whole field. Looking at the CMD pairs in Fig. 6 and Fig. 7, also at the contamination percentages, the reader can judge the level of contamination in each case.

MB 1 is located in the galactic plane ($b = -1°$). Despite its relatively large number of selected stars, 88% may be in the foreground, so the CMD has to be regarded with caution. NGC 1569 is located at $b = +11°$. With an expected contamination of about 10%, its CMD in Fig. 6 clearly shows a blue finger centered at $J - K_s = 1.2$, which is 0.4 mag redward of the blue finger of the CMD for the field stars in the Milky Way, suggesting that most of the stars bluer than $J - K_s = 1.0$ mag in the CMD of NGC 1569 in Fig. 6 are part of the Milky Way. Blue fingers like that in NGC 1569 have been evidenced previously in IC 10 by Borissova et al. (2000a). UGCA 92 sits at the same latitude $b = +11°$ and is a large target, therefore its contamination may be 70% and its CMD must be regarded with caution. The Orion dwarf is also situated at a similar low latitude ($b = -12°$), so its expected contamination may be as high as 90%. One of our faintest targets (see Fig. 1), DDO 47 is located at $b = +19°$. Its CMD looks very similar to that of the field; 90% of the stars may be in the foreground. Another well resolved galaxy, NGC 3738 is located at $b = +59°$ and is expected to suffer only 11% contamination. Its CMD looks different from its field pair, showing no sign of a blue finger. NGC 4163 and NGC 4190 also have high latitudes ($b = +78°$), and their blue fingers at $J - K_s = 1.0$ appear to be largely unaffected by foreground stars. NGC 5264 is located at $b = +32°$. This galaxy is relatively large, so about 26% of stars in the CMD may be contaminants, and its blue finger at $J - K_s = 0.8$ may be an artefact.

## 7. Surface Photometry

We have measured surface brightness profiles and total fluxes for the total flux coming from the unresolved component of each galaxy via the surface photometry package STSDAS/ELLIPSE. Working from images cleaned of stars (via KILLALL) instead of the original frames, our measurements have the huge advantage of being unaffected by surface brightness fluctuations due to inclusion of light from the resolved sources, which can be signifiant for bright stars.

Two galaxy frames observed at OAN-SPM (Cas 1 and NGC 1569) included more than 100 stars in their fields. Their resolved and unresolved components were separated using KILLALL. The other five OAN-SPM frames (UGC 4483, UGC 4998, UGC 5848, UGC 5979 and UGC 8508) showed very few stars, so we were unable to use KILLALL, as we were unable to construct reliable PSFs. For these five cases, we simply used IMEDIT to manually remove the stars in the frames.

### 7.1. The Unresolved Component

Unresolved images of the faintest galaxies were binned $5 \times 5$ pixels ($1'' \times 1''$) in order to improve signal-to-noise ratio. Following this process, 20 out of the 29 dwarfs observed at CFHT showed an unresolved component.

In order to measure surface brightness profiles, we used the ISOPHOT/ELLIPSE task on the cleaned $K_s$ images in two stages. First, we guessed the initial ellipse centres, ellipticities and position angles, but allowed them to vary freely with radius. Then we plotted the fits using the ISOPALL task and improved the ellipse parameters by analyzing the outer isophotes where the underlying old component is expected to set the geometry of the galaxy. Second, we fixed the centre, ellipticity, and the position angle and repeated the fitting process. By allowing only the surface brightness to vary with radius, it was possible to measure isophotes to fainter levels than in the first stage.

The upper graphs in the right panel beside each galaxy in Fig. 1 and Fig. 3 present the surface brightness profiles in $K_s$ and $J$ for the unresolved component of the galaxies observed at CFHT and OAN-SPM, respectively. The formal uncertainties in the surface photometry as given by ELLIPSE are plotted as error bars. Most of the errors are less than 0.1 mag arcsec$^{-2}$.

Table 1 gives the adopted values of the ellipticity ($e$) and major axis position angle ($PA$) from which the profiles were derived. Ellipticity errors listed by ELLIPSE are less than 0.05 in most cases, and position angles are uncertain to about 2 degrees. For the faint targets (DDO 47, DDO 167, DDO 187, UGC 4483, UGC 5979, UGC 6456, NGC 4789A), we estimate that errors reach 0.1 in ellipticity and 5-10 degrees in position angle.

### 7.2. Isophotal and Total Magnitudes

We define the *isophotal magnitude* ($m_I$) of each galaxy to be the total flux of the unresolved component. The isophotal magnitudes are given by the the instrumental detection limit as the total flux integrated to the faintest isophote. The detection limits were $\mu_{K_s} = 23$ mag arcsec$^{-2}$ and $\mu_J = 24$ mag arcsec$^{-2}$ for both facilities. We employed ISOPLOT to plot the magnitude growth curves, which at large radii converge asymptotically to the isophotal magnitudes. By adding the flux of the unresolved component to the flux of the resolved population (summed from all stars in the selected star catalog), we obtain the *total magnitude* ($m_T$). The isophotal and total magnitudes are listed in Table 3. The differences between the two are very small; less than 0.1 mag in most cases with a maximum of 0.2 mag in a few cases. This reveals the small contribution of the resolved sources to the total light from the galaxies, which will be discussed further in §8.1.

Only four galaxies from our sample were detected by 2MASS[7]. In Table 4, we compare measurements by listing the differences in the total magnitudes $\Delta m_J$ and $\Delta m_{K_s}$, ellipticity $\Delta e$, and position angle $\Delta PA$ (2MASS minus our values). We list the four targets in order of descending total luminosity. NGC 1569, the brightest galaxy, shows the smallest differences, with the 2MASS total magnitude being brighter by only $\sim$ 0.1 mag. The small difference may be due to imperfect removal of the bright foreground star 2MASS 04304867+6451504 ($K_s \sim 7.5$ mag) located just $\sim 1'$ North of the galaxy. The other three galaxies were fainter, and show increasing residuals in their total magnitudes, averaging about 1.5 mag and going in one case as high as 3 mag. In all three cases, 2MASS underestimates fluxes, likely due to the brighter detection limit.

---

[7]2MASS GATOR Catalog http://irsa.ipac.caltech.edu/applications/Gator

We observed NGC 1569 both at CFHT and OAN-SPM, allowing for a comparison between our two sets of results. The differences between the CFHT values and the OAN-SPM values are: $\Delta e = -0.05$, $\Delta PA = +5°$, $\Delta m_J = +0.03$ mag, and $\Delta m_{K_s} = +0.02$ mag. Agreement is outstanding.

A critical factor affecting photometry is the level of the sky. Our observing and reduction strategies removed sky to an accuracy of 0.02% in $K_s$ relative to the original signal, as we combined galaxy frames from which we subtracted appropriate sky frames close in time and position. Additionally, for the small galaxies, the sky level was checked by computing the median over the whole field, and then adjusted additively to zero. For the large galaxies, it was checked using the average of the medians in the corners. Thus, we believe that the uncertainties in the isophotal magnitudes are determined mainly by the zero points, whose typical errors are less than 0.1 mag (see §5).

Uncertainties in the total magnitudes include the errors in isophotal magnitudes (for the unresolved component) and the formal errors in fluxes in the selected star catalogs. The former are about 0.1 mag, as specified above. The later mostly cancel out statistically, due to the large number of stars detected in each galaxy.

Throughout this paper we have not corrected our magnitudes and colours for extinction and reddening, which are expected to be less than 0.05 mag for most of the dIs but which may amount to 0.38 mag extinction for MB 1 which is at $b = -1°$. We plan to examine these effects in a future paper which will enlarge our dI sample.

### 7.3. Astrometry

We measured the ICRS/J2000.0 absolute positions of the galaxy centres, which were derived using the unresolved galaxy images. For all galaxies except four, the reference stars were exclusively 2MASS catalog stars. For the four remaining frames having only one or two 2MASS stars in the FOV, we completed the reference system with a few GSC 2.2 or USNO-B1 positions. The absolute J2000 positions are listed in Table 1.

Two sources are responsible for the uncertainties in galaxy positions listed in Table 1: the reference system, and the ellipse centering algorithm. The astrometric accuracy of the 2MASS Point Source Catalog for stars with magnitudes in the range of $9 \leq K_s \leq 14$ is about $0.07 - 0.08''$, reaching $0.2 - 0.3''$ for its faintest sources [8]. The 2MASS observing epoch (1997-2001) is very close to our runs (2001-2004), so proper motions of the faint stars used for our reference fields do not affect our positions by more than $\sim 0.1''$.

---

[8] Cutri et al., 2003, User's Guide to the 2MASS All-Sky Data Release, http://www.ipac.caltech.edu/2mass/releases/docs.html

As described previously, the fitting of the unresolved component was done in two steps, in the first allowing the centre, ellipticity, and the position angle of the isophotes to be free parameters, while in the second fixing them at values found for the outer regions. In the second step, the errors in the centres came out to be less than $\sim 0.5''$. Therefore we can adopt $1''$ as the maximum error in astrometry for most of our targets.

### 7.4. Profile Fitting

James (1991, 1994) pioneered NIR surface photometry ($J$, $H$, and $K_s$) of dwarf galaxies. He modeled surface brightness profiles (SBPs) with an exponential law for dIs and BCDs and a de Vaucouleurs $r^{1/4}$ law for dEs. Bergvall et al. (1999) employed a pure exponential profile in order to fit the disks of a sample of low surface brightness blue galaxies observed in $J$, $H$ and $K_s$. Patterson & Thuan (1996) used an exponential law with a limiting inner limit in radius in order to fit the underlying component of a sample of 51 dwarfs (mostly dIs) observed in the visible.

A combination of two or three different models has been employed by some authors to improve the fits to the SBPs. Gavazzi, et al. (2000) used the exponential law, the de Vaucouleurs law, or a combination of the two in order to fit the light profiles for dIs and BCDs observed in $H$ and $K_s$ in five nearby clusters. Gavazzi, et al. (2001) employed a de Vaucouleurs law, an exponential law, a "mixed" profile (sum of the two), and a "truncated" profile (a blend of the two defined by a critical radius). This latter work was based on $H$-band observations of 75 faint galaxies, mostly dwarf ellipticals (dEs) in the Virgo cluster. Using observations in the visible of 12 BCDs and 2 other bright starburst galaxies, Papaderos et al. (1996) decomposed SBPs into three components: an exponential at large radii, a plateau at intermediate radii, and a Gaussian at small radii.

As mentioned previously, we derived our SBPs using the ELLIPSE task of the ISOPHOT package in IRAF. Most of the SBPs show a flattened core and exponential component extending down to a surface brightness of about $\mu_{K_s} = 23$ mag arcsec$^{-2}$ and $\mu_J = 24$ mag arcsec$^{-2}$, as can be seen in Fig. 1 and Fig. 3.

We attempted to fit our SBPs over the entire length of the semimajor axes using an exponential or a plateau component, either alone or as a sum. These attempts failed to provide good fits at both small and large radii. Given that we are especially interested in properly characterizing the outer regions, where we believe that the old stellar population dominates, we looked at a variety of other functions to simultaneously fit both the inner and outer regions. Good fits could be achieved for most galaxies using the following hyperbolic secant function:

$$I = I_0 \operatorname{sech}(r/r_0) = \frac{I_0}{\cosh(r/r_0)} = \frac{2I_0}{e^{r/r_0} + e^{-r/r_0}} \tag{5}$$

Here $I$ represents the fitted flux at radius $r$, defined as the distance from the centre along the semimajor axis. $I_0$ is the *central intensity* (expressed in counts/pixel), and $r_0$ represents the *scale length* of the profile (expressed in pixels).

A hyperbolic secant was previously employed by van der Kruit & Searle (1981). The authors used the function *sech²* to model the vertical distribution of light in galactic disks based upon dynamical expectations for an isothermal disk.

In magnitude units, the fit is given by the following equation:

$$\mu = zp_s - 2.5\log(I) = zp_s - 2.5\log\left(I_0 \operatorname{sech}(r/r_0)\right) = zp_s - 2.5\log\frac{I_0}{\cosh(r/r_0)} \tag{6}$$

where $zp_s$ represents the zero-point of the surface brightness magnitude system. At larger radii, equation (6) is convergent to the *sech magnitude* ($m_S$), while curving and leveling out at near-zero radii toward the *central surface brightness* ($\mu_0$). Expressed in mag arcsec$^{-2}$, the central surface brightness is simply:

$$\mu_0 = zp_s - 2.5\log(I_0) \tag{7}$$

To perform the actual fitting of the SBPs generated by ELLIPSE, we employed the task NFIT1D (in the FITTING package of STSDAS), entering Equation (5) as our user specified function for the USERPAR/FUNCTION parameter. For plotting purposes, fitted fluxes were converted into mag arcsec$^{-2}$ using the zero-points for the frames. We have plotted the fits on top of the SBPs in the upper graphs in the right panels of Fig. 1 and Fig. 3.

In a few cases (NGC 1569, NGC 3738, UGC 5848, UGC 5979), the SBPs show some excess flux at small radii. This could be evidence of starburst activity near the galaxy centre, which has been previously suggested for NGC 1569 (Greggio, et al. 1998; Aloisi et al. 2001) and NGC 3738 (Hunter, Gallagher & Rautenkranz 1980; Karachentsev et al. 2003b). A few other cases (Mark 178, NGC 3741, Cassiopeia 1) present some flux deficit at small radii. Where necessary, to prevent results from being biased by flux excesses or deficits at small radii, we restricted fits to the outer regions. For those galaxies where this was done, we plotted the fitting interval as a line in the upper part of the graphs in Fig. 1 and Fig. 3. Table 3 lists the central surface brightness $\mu_0$ (in mag arcsec$^{-2}$) and the scale length $r_0$ (in arcsec) in both $J$ and $K_s$. For most galaxies, the scale lengths in $J$ and $K_s$ agree to within about 1 arcsec.

### 7.5. Sech Magnitudes, Semimajor Radius

The "sech magnitude" of the fit to the unresolved component of each galaxy can be calculated from Equation (5). Let $a$ and $b$ be the semimajor and semiminor axis of the isophotes, $e$ the ellipticity, and $I(a)$ the average surface brightness in an elliptical annulus with area $dA$. Then,

$$dA = d(\pi ab) = d(\pi a^2(1-e)) = 2\pi a(1-e)da \qquad (8)$$

and the total flux $F$ is given by:

$$F = \int_0^\infty I(a)dA = 2\pi(1-e)I_0 \int_0^\infty a \operatorname{sech}(a/R_0)da \qquad (9)$$

By substituting $x = a/r_0$ in the last integral, the integrated (sech) flux becomes:

$$F = 2\pi I_0 r_0^2 (1-e) \int_0^\infty x \operatorname{sech} x\, dx \qquad (10)$$

The integral on the right side can not be resolved analytically. Using MAPLE, we obtained a numerical approximation of 1.83193119. With this result, the *sech magnitudes* can be calculated from:

$$m_S = zp_s - 2.5 \log(11.51036 I_0 r_0^2 (1-e)) \qquad (11)$$

with $zp_s$ representing the frame zero-point for point sources.

We include in Table 3 the *sech magnitudes* ($m_S$) for the unresolved component of the galaxies as calculated from Equation (11). Most sech magnitudes are very close to the isophotal magnitudes (in most cases within 0.2 mag). In only four cases the discrepancies between $m_S$ and $m_I$ reach 0.4 mag, due to the faintness of the unresolved component or the presence of a starburst (NGC 1569). Fig. 9 plots the difference between the isophotal magnitude $m_I$ minus the sech magnitude $m_S$ as a function of the $m_I$.

Using Equation (6), we have calculated the semimajor axis $r_{22}$ corresponding to the $\mu_{K_s} = 22$ as well as $r_{23}$ corresponding to $\mu_J = 23$ mag arcsec$^{-2}$. The positive solution for the 2nd order equation in $r_{22}$ comes to

$$r_{22} = r_0 \ln \frac{1 + \sqrt{1 - A_{22}^2}}{A_{22}} \qquad (12)$$

where $A_{22}$ is given by

$$\log A_{22} = \frac{zp_s - 22 - 2.5 \log I_0}{2.5} \qquad (13)$$

There are similar formulae for $r_{23}$. Our values for $r_{22}$ and $r_{23}$ are listed in Table 3.

Based on the NFIT1D rms error analysis, we evaluate typical uncertainties in $r_0$, $r_{22}$ and $r_{23}$ to be about 2%, and the uncertainties in the central surface brightness $\mu_0$ and the sech magnitude $m_S$ to be less than $\sim 0.1$ mag arcsec$^{-2}$ and $\sim 0.1$ mag, respectively. From a purely empirical point of view, the estimated uncertainty of 0.1 mag in $m_S$ agrees resonably well for most of the galaxies plotted in Fig. 9 which show $m_I - m_S \lesssim 0.1$.

### 7.6. Color Profiles

Color profiles for the underlying unresolved component of each galaxy can be plotted by combining $J$ and $K_s$ surface brightnesses via the TCALC task of TABLES. For each galaxy the color profile is given in Fig. 1 and Fig. 3 as the bottom graph of the right panel beside the picture. Most galaxies show a remarkably constant $J - K_s = 0.8$ to 1.0 mag, close to the color of the blue finger seen in the CMDs.

## 8. NIR Properties of dIs

### 8.1. Resolved / Total Flux Ratios

The brightest stars in dwarfs are relatively young, so the ratio of the resolved flux to the total flux gives some idea of how well the total light reflects the mass of the old population.

We define the *resolved flux* as the sum of the fluxes of the stars detected by KILLALL down to $M_{K_s} = -7.5$. We deliberately do not include all resolved sources in the resolved flux since our detection limit for resolved sources varies from one object to another, based on their distances. Indeed, we include only those sources brighter than $M_K = -7.5$ mag since our observations were tuned to detect such sources in all galaxies. The *total flux* is defined as the flux of the unresolved component plus the sum of the fluxes of all resolved sources (faint stars), regardless of brightness. The ratio of the resolved to the total flux is denoted here as "Resolved/Total".

In order to check the sensitivity of the results to resolution, we degraded the resolution of GR 8 by a factor of two, as if its distance were moved from 2.0 to 4.0 Mpc. We repeated the star detection process and recalculated the resolved/total ratio; it agreed with the first to within 0.2%. Also, from our entire sample we see no trend in the resolved/total ratio with distance. We conclude that over the range of distance sampled here, results for the flux ratios are not affected by variations in resolution.

The flux ratios for the resolved galaxies observed at CFHT are reported in Table 3 (in per cent). No ratios are given for the galaxies observed at OAN-SPM, as we did not quantify the resolved flux due to the lower spatial resolution. The resolved fluxes for all galaxies are very small in comparison with the total fluxes; in most cases, the ratio is less than 5%. In fact, the true ratios are even lower given that the resolved population is contaminated to varying degrees by foreground stars. Ratios in $J$ are 1.5–2 times larger than in $K_s$. Thus, the total NIR flux in a dI appears to be contributed mainly by stars with $M_{K_s} \gtrsim -7.5$, with $K_s$ being less sensitive than $J$ to the population of luminous stars.

One can employ population syntheses (Girardi, et al. 2000, 2002) to evaluate the fraction of the flux from a burst of star formation contributed by stars brighter than $M_{K_s} = -7.5$. Although we don't know the detailed history of star formation in the dwarfs, we can look at how the proportion of stars brighter than $M_{K_s} = -7.5$ for any given burst depends upon when the burst occurred, and thereby judge what our "old population" really represents. We performed this analysis using a Salpeter initial mass function, $N(M) = AM^{-2.35}$, and isochrones for metallicity Z=0.0004, which is a typical low metallicity dI. Fig. 10 plots the flux fraction as a function of stellar age. The peak of the distribution is around $0.2 - 0.3$ Gyr, with almost all the flux in a burst contributed by stars brighter than $M_{K_s} = -7.5$ mag. The proportion is above 50% for bursts younger than about 3 Gyr. For bursts older than about 6 Gyr, there are no stars brighter than $M_{K_s} = -7.5$ mag. Since 3 Gys is a very generous definition of "recent" star formation, it is clear that the resolved fluxes represent the majority of the flux from such stellar populations. Given that the resolved flux is also a small fraction of the total flux, the remaining flux is a reliable measure of the flux from the underlying galaxy, unperturbed by recent star formation.

The appropriate age to associate with the unresolved component depends upon how much the light of this component is contaminated by unresolved stars which formed along side the resolved component. A constraint can be gained from the fraction of the light which is resolved. Suppose all resolved stars came from a single burst. Define $R_b(t)$ to be the flux from resolved burst stars, i.e., those brighter than $M_K = -7.5$ mag, and $U_b(t)$ to be the flux from unresolved burst stars, i.e., stars fainter than $M_K = -7.5$ mag, both at time $t$ since the burst. Then the fraction of burst stars resolved at time $t$ is $k(t) = R_b(t)/[R_b(t) + U_b(t)]$. Approximate the light of all stars formed prior to the burst to be unresolved, and call the flux $U_{old}$. Then the fraction of the total light of all stars ever formed which are observed to be brighter than $M_K = -7.5$ mag is $f = R_b(t)/[R_b(t) + U_b(t) + U_{old}]$. It can be shown that

$$k(t) = \frac{f + fU_{old}/U_b}{1 + fU_{old}/U_b} \quad (14)$$

Given a constraint on $U_b/U_{old}$, and an observation of $f$, it is possible to solve for $k$. With the aid of Figure 10, the age $t$ of the burst can be estimated. If we require that less than 10% of the old component be burst stars, then the observation that $f = 0.05$ requires that the burst have happened in the last 4 Gyr. Even though the actual star formation history may be more complicated, this analysis constrains the age of the unresolved component in dIs to be greater than 4 Gyr.

### 8.2. Correlations

Here we examine correlations between different properties of the dIs listed in Table 3. To express sizes, we employ the semimajor axis $r_{22}$ for $K_s$, $r_{23}$ for $J$ and scale length $r_0$ for $K_s$. For the absolute magnitudes, we use mostly the sech magnitude $M_S$, while for comparison the isophotal magnitude $M_I$ and the total magnitude $M_T$. To characterize the centres, we employ the central surface brightness $\mu_0$. To characterize motions, we use $W_{20}$, the width of the HI line profile, measured in km/s at 20% of the peak (included in Table 1). We focus on the $K_s$ band, because it quantifies better the old population, but similar results are obtained in $J$. In each graph to be discussed below, typical uncertainties in each parameter are plotted as an error cross.

#### 8.2.1. Size and Brightness

The size of a dI is expected to correlate with the luminosity of the old component, which is responsible for most of the baryonic mass. A linear trend between the logarithm of the radius and the absolute magnitude was seen previously in the $B$ band by Hidalgo-Gamez and Olofsson (1998), who studied a sample including 176 irregular galaxies (spiral/irregulars and irregulars). Comparing BCDs with dIs, Papaderos et al. (1996) found a clear linear trend of the scale length as a function of the absolute magnitude in $B$, with BCDs showing a steeper correlation than dIs. In the NIR, a similar correlation has been found between the effective radius and luminosity in $H$, based on a larger sample of different types of galaxies, but the scatter is pretty large (Scodeggio et al. 2002).

Fig. 11 and 12 show, respectively, the relation between the scale length $r_0$ and the semimajor axis $r_{22}$ (both in kpc), as a function of the sech magnitude, $M_S$. Both plots show good correlations; of the 27 galaxies having data in our sample, only four points appear to reside outside the linear trends: UGC 5979, UGC 5848 (our most distant targets, both having uncertain distances), Holmberg IV (one of our largest targets), and NGC 1569 (well known for its starburst). Rejecting the four deviant points, a linear fit in $K_s$ gives the following solutions:

$$r_0 = (-0.81 \pm 0.24) + (-0.07 \pm 0.01)M_S \qquad (15)$$

$$r_{22} = (-5.34 \pm 0.40) + (-0.38 \pm 0.02)M_S \qquad (16)$$

We mention here that the correlation between the semimajor radius and the sech magnitude is not artificial (cf. Eq. 11), as we found a very similar trend replacing the sech with the total magnitude. Based on Equation (10), the absolute magnitude is expected to be a linear combination of the central surface brightness and the logarithm of the scale length, i.e.:

$$M_S = k + \mu_{0S} - 5\log r_0 \qquad (17)$$

where k is a constant. The thin curve in Fig. 11 represents a curve of constant surface brightness, $\mu_{0K} = -19$ mag arcsec$^{-2}$. Clearly, the dwarf sequence is one of varying central surface brightness.

The uncertainties in $r_0$, $r_{22}$, and $r_{23}$ are determined mostly by the errors in the galaxy distances, because formal fitting errors are small (about 2%). The tip of the red giant branch (employed in 20 cases) gives an error of about 10%. The brightest blue stars (used in 9 cases) yield a scatter as large as 50% in distance. Hubble distances (employed only in two cases, UGC 5848 and UGC 5979) are supposed to be good to about 12%, although over the distance range covered here, local flow corrections are significant.

The correlations confirm that brighter dIs are larger. The fact that $r_0$ grows with luminosity also reveals that bright dIs have a larger central plateau. Correlations with the sech magnitudes are tighter than with isophotal and total magnitudes, in the later case confirming the closer connection to the underlying mass.

### 8.2.2. Central Surface Brightness

In $B$, Hidalgo-Gamez and Olofsson (1998) did not find any correlation between the surface brightness and the absolute magnitude, based on a sample of 176 irregular galaxies.

Using a larger sample observed also in $B$, Papaderos et al. (1996) studied the central surface brightness of the underlying exponential component as a function of the total luminosity. Notwithstanding the large scatter, there was a suggestion that the central surface brightness grows with increasing luminosity for dwarf galaxies other than BCDs.

Fig. 13 shows the central brightness $\mu_0$ as a function of the sech magnitude $M_S$. Rejecting the four outliers discussed above, there is a clear trend between the two parameters, in the sense that more luminous galaxies have brighter cores. Probably this has not previously been seen so clearly because of failure to separate young and old components. Moreover, rejecting NGC 3738 (which also hosts a burst), the correlation appears to level off at high luminosities, suggesting an upper limit to the central density of mass.

From Equation (17), the absolute magnitude is expected to depend linearily on the central surface brightness. The thin line in Fig. 13 represents the line of constant scale length, $r_{0K} = 0.4$ kpc.

A plot of the central surface brightness versus the Resolved/Total ratios does not show any correlation. This suggests that the fraction of the light which is young is not directly linked to the density in the core.

No correlation has been found between the central surface brightness and the scale length. This shows that both variables are required to define the surface brightness profile and luminosity.

### 8.2.3. Colors

In the $B$ band, Hidalgo-Gamez and Olofsson (1998) found no relation between their colours and absolute magnitudes for a subsample including 89 irregular galaxies. Employing a larger less homogenous dwarf sample, Scodeggio et al. (2002) also did not find a correlation between $B - H$ colors and the logarithm of $H$ luminosity, although for spirals a linear trend is visible.

As Fig. 14 shows, our data reveal a correlation between colors and total magnitudes, with brighter galaxies being redder. Similar trends are obtained using the isophotal and sech magnitudes. Only a few points stand outside the trend, including UGCA 92, UGC 4483, and DDO 47 (all very faint), Ho IV (larger than the field), and MB 1 (suffering high reddening). The cause of the variation with magnitude is unlikely to be the star formation rate because the central surface brightness and color does not depend on the fraction of light which is resolved. Also, across any given dwarf, the color does not vary much with radius (see § 7.6).

Rather, the reddening is more likely to be a manifestation of increasing metallicity. Following previous research in the visible (e.g., Richer & McCall 1995), the metallicity is expected to correlate with the NIR luminosity, with brighter galaxies having higher metallicities. The colors in Figure 14 imply that the light is entirely dominated by evolved stars (Girardi, et al. 2000, 2002). Galaz, et al. (2002) showed that the $J - K_s$ color index is clearly very metal-sensitive and relatively age-insensitive. Assuming a single stellar metallicity, the $J - K_s$ colours can vary by at most 0.15 mag as a result of differing star formation histories. This is about 3 times less than the color range which we see in Figure 14. On the other hand, a variation by a factor of 15 in metallicity is expected to lead to about 0.3 mag variation in color, which is much closer to the range observed.

### 8.2.4. The Tully-Fisher Relation

For luminous disk galaxies, the absolute magnitudes correlate with HI line-width, i.e., rotational velocities (Tully and Fisher 1977). However, it is not clear how well dIs, known to rotate slowly, follow the Tully-Fisher correlation, because it is unknown how random motions scale with mass and how much star formation perturbs luminosities.

In a study of 50 giant and dwarf galaxies selected from the Virgo Cluster Catalogue, Pierini & Tuffs (1999) presented evidence that luminous dIs follow the Tully-Fisher relation in $K'$ but their sample only went as faint as $M_K = -17$. Of our field dI sample, 22 galaxies have $W_{20}$ measurements listed in the RC3 (de Vaucouleurs et al. 1991)[9]. Fig. 15 plots with filled circles their sech absolute magnitudes versus log$W_{20}$. On the same diagram, 16 dwarfs from Pierini & Tuffs (1999) are plotted with open circles. The absolute magnitudes were calculated from the apparent $K'$ magnitudes taken from the quoted paper using an updated Virgo distance modulus of 30.62 mag anchored to the maser distance to NGC 4258 (Freedman et al. 2001; see the Appendix). For these galaxies, too, we employed $W_{20}$ from the RC3.

As one can see from Fig. 15, $M_K$ correlates with $W_{20}$ down to $M_K = -13.5$, albeit with a standard deviation of 0.98 mag and a correlation coefficient of only 0.57. The most deviant point in our sample is NGC 5264, possibly because of an uncertain distance based on its radial velocity. A linear fit to our galaxies, excluding NGC 5264, gives a slope of $-6.4 \pm 1.3$. The fit is shown as a solid line in Fig. 15. Our data overlap the faint end of the NIR diagram of Pierini & Tuffs (1999). Fitting spirals and dwarfs together, these authors found a slope of $-9.7 \pm 0.3$. The dashed line in Fig. 15 shows their slope. Although the error quoted by Pierini & Tuffs (1999) suggest a more solid fit than ours, their fit is biased toward spirals.

---

[9] Accessed with VizieR, online at CDS, http://vizier.u-strasbg.fr/viz-bin/VizieR

Their points for dwarfs look more scattered than ours. Both sets of data show a larger spread at the bright end, possibly because of enhanced rotational motions and the varying angles at which they are viewed. However, little change is observed when $W_{20}$ is corrected for tilt on the basis of isophotal axis ratios.

### 8.2.5. A Fundamental Plane for dIs

The large scatter in the Tully-Fisher relation hinted that dI luminosities depend upon more than just internal motions. Thus, we investigated whether residuals might be linked to the central surface brightness $\mu_0$ or the scale length $r_0$.

Fig. 16 shows that the absolute sech magnitude is strongly correlated with the linear combination of $\log(W_{20})$ and the central surface brightness, $\mu_0$. The three deviant points are UGC 5979 (whose distance is approximate, determined from the Hubble law), Holmberg IV (a galaxy larger than the field of view, also having an approximate distance determined using the three brightest blue stars), and NGC 5264 (which might be classified as a "dwarf spiral", e.g., Lee, et al. 2003, in which case tilt corrections to $W_{20}$ should be applied). Rejecting these three outliers, the fit can be expressed through the following correlation:

$$M_S = (-4.27 \pm 0.61)\log(W_{20}) + (0.70 \pm 0.09)\mu_0 - (22.08 \pm 2.40) \qquad (18)$$

With a rms deviation of 0.43 mag and a correlation coefficient of 0.91, we think this may provide a useful avenue for determining distances to dIs.

We compared $W_{20}$ with $W_{50}$ (Karachentsev et al. 2004), and found that $W_{20}$ produced a stronger correlationn. We also tried correcting for inclination, computing $W_{20}/(2\sin i)$ using the algorithm of Staveley-Smith, Davies & Kinman (1992) along with the ellipticities listed in Table 1. However, the improvement was very small, suggesting that internal motions in the majority of our dwarfs are predominantly random. This is a major advantage over spirals in the sense that good distances can be determined without correcting $W_{20}$ for tilt.

Denoting the mean surface brightness by <SB>, Equation 18 implies that $L_K \propto \sigma^{1.71}$ <SB>$^{0.70}$, where $L_K$ is the total luminosity, $\sigma$ the central velocity dispersion. For elliptical galaxies, Djorgovski and Davis (1987) found $L \propto \sigma^{3.45}$ <SB>$^{-0.86}$ in the Gunn $r$ band. Dwarf irregulars are distinct from ellipticals in that luminosity increases with surface brightness.

## 9. Conclusions

Nearby dwarf irregular galaxies (dIs) and blue compact dwarfs (BCDs) are important probes for studying the formation and evolution of galaxies. In an effort to extract information about old stellar populations, a sample of 34 dIs in the field, most of them located closer than 5 Mpc and having known or soon-to-be known metallicities, have been imaged using NIR arrays at CFHT in Hawaii and the OAN-SPM in Mexico, between 2001-2004. Besides detecting the unresolved component in 25 of 34 targets, which is primarily associated with the old stellar population, we also resolved stars as faint as $M_{K_s}$ = -7.5 mag out to 5 Mpc.

By separating the resolved sources associated with each galaxy from the unresolved component, we were able to determine for the first time the contribution from the resolved stellar component to the total light in the NIR. In nearly all galaxies, the resolved population brighter than $M_{K_s} = -7.5$ mag represents less than 5% of the flux in $K_s$ and 7 to 10% in $J$. Population syntheses reveal that stars brighter than $M_{K_s} = -7.5$ contribute more than 50% of the light from bursts as far back as 3 billion light years. Given the small contribution of the resolved flux to the total flux of dIs, together with the constant NIR colours, one can regard NIR fluxes for dIs as approximating the light from a population older than about 4 Gyr, and thereby use them as gauges of stellar mass.

Surface brightness profiles of the unresolved components have been successfully fitted across the whole range of radii with a hyperbolic secant function defined by only two parameters: the central surface brightness $\mu_0$ and the scale length $r_0$. Only two cases (NGC 1569 and NGC 3738) showed flux excesses in the centre, but both are known to host central starbursts similar to those seen in BCDs. These two galaxies will be addressed in a future paper devoted to BCDs (Vaduvescu, Richer and McCall, 2005, in preparation). Isophotal, total and sech NIR magnitudes have been calculated for all galaxies showing an unresolved component, along with semimajor axes at $\mu_J = 23$ mag arcsec$^{-2}$ and $\mu_{K_s} = 22$ mag arcsec$^{-2}$.

We searched for correlations between galaxy size, absolute magnitude (sech, isophotal, and total), central surface brightness, color, and the Resolved/Total ratio. Good linear correlations were found between the scale length and the sech magnitude, and between the semimajor axis and the sech magnitude. Also, correlations were found between colors and isophotal magnitudes, and between the central surface brightness and the sech magnitude. Overall, galaxies with more luminous old components are larger, redder and brighter in the centre. Thus, size, color, and the extent of the central plateau appear to be influenced by the mass of the old component. Redder colors may be a consequence of increasing metallicity, because the fraction of the light which is young shows no trend with absolute magnitude or central surface brightness.

At the lower luminosity end of our dI sample, the Tully-Fisher relation shows considerable scatter in $K_s$. The scatter appears to be tied to variations in surface brightness. A new "fundamental plane" is found to relate the sech absolute magnitude, the central surface brightness, and the hydrogen line-width uncorrected for tilt. The residuals are low enough (0.4 mag) that it offers considerable potential as a distance indicator for star-forming dwarfs.

For 29 galaxies, color-magnitude diagrams for the resolved component have been derived. Three CMDs include more than 1000 stars in both $K_s$ and $J$, while another 15 CMDs have more than 100 stars. Most of the CMDs show a main blue finger centered around $J - K_s = 1$ mag. In some cases, a red tail extends from the finger out to $J - K_s = 2.5$ mag. The color profiles of the unresolved components show a remarkably constant $J - K_s$ =0.8 to 1.0 mag, which is close to the color of the main finger in the CMDs.

We thank the CFHT and OAN-SPM time allocation committees for granting us the opportunity to observe. CFHT is operated by the National Research Council of Canada, the Centre National de la Recherche Scientific, and the University of Hawaii. Thanks are conveyed to the National Research Council of Canada for funding the observing expenses of OV. MLM is grateful to the Natural Sciences and Engineering Research Council of Canada for its continuing support. MGR acknowledges financial support from CONACyT grants 37214-E and 43121 and DGAPA grant IN112103. MGR thanks G. Garcia, G. Melgoza, F. Montalvo, and S. Monrroy for their help with the observations. For our data reductions, we used IRAF, distributed by the National Optical Astronomy Observatories, which are operated by the Association of Universities for Research in Astronomy, Inc., under cooperative agreement with the National Science Foundation.

## A. APPENDIX

The majority of the distances adopted in this paper have been derived using modern photometric data with either primary or secondary distance indicators. All distances have been homogenized to a common distance scale anchored by the maser distance to NGC 4258 (Herrnstein et al. 1999). Furthermore, the distance estimates have been homogenized to a common extinction scale. Specifically, all but one of the Galactic extinction estimates originate from the reddenings of Schlegel, Finkbeiner and Davis (1998) (hereafter referred to as SFD). The reddening of MB1 has been assumed to be identical to that of Maffei 1 (Fingerhut 2003). The reddenings have been converted to broadband extinctions using the York Extinction Solver (YES)[10]. Given an SFD reddening for a target, YES computes the corresponding optical depth at 1 micron ($\tau_1$), using an elliptical SED (upon which SFD reddenings are based), then converts this value into broadband extinctions in the desired filters. By employing YES to calculate broadband extinctions, using a SED representative of a target appropriately redshifted and extinguished, we avoid source-dependent shifts in the effective wavelengths of broadband filters. Further details can be found in McCall (2004).

The homogenized distance estimates are listed in Table 1 along with the distance method, original distance reference, and $\tau_1$. The homogenization process for each distance method is described briefly in the following paragraphs. Further details will be provided in a forthcoming paper by Fingerhut et al. (2005, in preparation).

### A.1. Tip of the Red Giant Branch (TRGB)

Twenty dIs in our sample have distances determined from the absolute magnitude of the TRGB, which marks helium ignition in the degenerate core of low-mass stars. In the $I$-band, the TRGB has been found to be nearly invariant over a wide range of metallicities (see, e.g., Lee, Freedman and Madore 1993), thereby serving as a powerful standard candle. We adopt the value of $M_{I,\mathrm{TRGB}} = -4.06 \pm .07\,\mathrm{mag}$ (Ferraresse et al. 2000), which is calibrated by galaxies with Cepheid distances anchored to an LMC distance of $\mu_{\mathrm{LMC}} = 18.5 \pm 0.13\,\mathrm{mag}$. The NGC 4258 distance scale brings the LMC distance closer by 0.19 mag, inducing an increase in $M_{I,\mathrm{TRGB}}$ by this same amount. The modification to $M_{I,\mathrm{TRGB}}$ has no effect on the uncertainty in this quantity due to the equivalent error bars on the LMC distance in both systems. Furthermore, $M_{I,\mathrm{TRGB}}$ was derived from $I$-band photometry corrected for Galactic extinction using SFD reddenings, but with a slightly larger value of $A_I/E(B-V)$ than the average of the values computed by YES for the galaxies used to calibrate the TRGB. However, at the low reddenings of the TRGB calibrators, the different values of $A_I/E(B-V)$ have a negligible effect on $A_I$. As a result, $M_{I,\mathrm{TRGB}}$ does not need to be shifted to our extinction scale. We therefore adopt $M_{I,\mathrm{TRGB}} = -3.87 \pm 0.07\,\mathrm{mag}$.

---

[10]The York Extinction Solver (YES) is a web-based application developed in the Department of Physics and Astronomy, York University and hosted by the Canadian Astronomy Data Centre (CADC). It can be accessed at http://cadcwww.hia.nrc.ca/astrocat/yes

The apparent magnitudes of the TRGB for the dIs in our sample have been corrected for extinction using the Galactic extinction as a reasonable estimate of the total extinction. This has been justified by Lee et al. (2005, in preparation), who found that the total extinction of dIs as estimated from the Balmer decrement in HII regions is equivalent, within errors, to the Galactic extinction determined from independent methods.

The dominant sources of uncertainty in the TRGB distances are the uncertainty in $M_{I,\mathrm{TRGB}}$ and the uncertainty in the apparent $I$-band magnitude of the TRGB. For the galaxies in our sample with TRGB distances, the average uncertainty in the latter quantity is 0.17 mag. Hence, the average error associated with the TRGB distance moduli is 0.18 mag.

## A.2. Brightest Blue Stars (3BS)

Nine galaxies have distances determined from the average $B$ magnitude of the three brightest stars. The most recent calibration of this method is that of Karachentsev and Tikhonov (1994) based on a sample of 43 galaxies. Improved distance estimates and Galactic extinctions are now available for a large part of this sample, so we have recalibrated the 3BS method using the 29 galaxies which now have either TRGB or Cepheid distances. After homogenizing the TRGB and Cepheid distances as well as the extinction estimates to our adopted scale, we obtain a new 3BS calibration of:

$$M_{B3} = -0.49(B3_0 - B_0^T) - 3.90 \quad (A1)$$

where $B3_0$ is the average magnitude of the 3 brightest stars in $B$ corrected for extinction, $M_{B3}$ is the absolute magnitude corresponding to $B3_0$, and $B_0^T$ is the total magnitude of the galaxy in $B$ corrected for extinction. Equation A1 has an RMS scatter of 0.97 mag. Further details of this calibration will be provided in a future paper. As with the TRGB distances, the 3BS distances have been derived using the Galactic extinction as an estimate of the total extinction of dIs.

### A.3. Surface Brightness Fluctuations (SBF)

For UGC 4998, the only modern distance estimate is the value obtained by Jerjen et al. (2001) from the SBF method in the $R$-band. Using TRGB distances for a sample of dE galaxies, they construct a semiempirical calibration for the SBF method as a function of $B - R$ colour. Jerjen et al. (2001) judge the uncertainty in distances derived from their calibration to be $\sim 10\%$. Because the zero-point of the calibration is set by their adopted TRGB of $M_{I,\mathrm{TRGB}} = -4.05\,\mathrm{mag}$, we shift their calibration by an additive constant of 0.18 mag to anchor it to our adopted value of $M_{I,\mathrm{TRGB}}$. The galaxies used to calibrate the relation were corrected for Galactic extinction using the reddenings of SFD. The reddenings are too low to have led to an appreciable difference in broadband extinctions compared with those computed by YES. Hence, the SBF calibration is already anchored to our adopted extinction scale and, with the shift of 0.18 mag, is sufficient for yielding distances homogenized to the rest of our sample.

### A.4. Hubble Distances (HUB)

Hubble distances have been adopted for the two most distant galaxies owing to their lack of suitable photometry. The Hubble constant is $79.2 \pm 9.9\,\mathrm{km\,s^{-1}\,Mpc^{-1}}$ based upon the Cepheid calibration of Freedman et al. (2001), but re-anchored to the maser distance of NGC 4258. The heliocentric radial velocities were corrected for the solar motion toward the Local Group centroid using the solar apex vector determined by Courteau and van den Bergh (1999). The velocities were then further corrected for their infall toward the Virgo Cluster and the Great Attractor using the linear multiattractor model formulated by Marioni et al. (1998). The uncertainty in the radial velocities of both galaxies is less than 1%. The uncertainty in the Hubble distances is therefore equal to the 12% uncertainty in the Hubble constant.

### A.5. Membership Distances (MEM)

The galaxies Cas 1 and MB 1 lack suitable data for a photometric distance estimate. However, both galaxies are believed to be associated with the nearby IC 342/Maffei group of galaxies. Recently, the distance to this group was found to be $3.3 \pm 0.2/$, Mpc, which is the weighted average distance of its three dominant members (Fingerhut 2002, 2003). We assign this distance to Cas 1 and MB 1.

Table 1. dI Sample

| Galaxy (1) | $\alpha$ (J2000) (2) | $\delta$ (J2000) (3) | $b$ (4) | $e$ (5) | $PA$ (6) | $\tau_1$ (7) | $\mu$ (8) | Method (9) | Ref (10) | $W_{20}$ (11) |
|---|---|---|---|---|---|---|---|---|---|---|
| Cassiopeia 1 | 02:06:05.4 | +69:00:12 | +7 | 0.00 | 0 | 1.18 | 27.59 | MEM | fin02 | – |
| MB 1 | 02:35:35.1 | +59:22:42 | −1 | 0.60 | −66 | 1.13 | 27.59 | MEM | fin02 | – |
| Cam A | 04:25:16.3 | +72:48:21 | +16 | (a) | (a) | 0.25 | 26.28 | TRGB | kma99 | (a) |
| NGC 1569 | 04:30:49.3 | +64:50:55 | +11 | 0.45 | −64 | 0.80 | 26.37 | TRGB | mk03 | 84 |
| UGCA 92 | 04:32:03.5 | +63:36:58 | +11 | 0.50 | +51 | 0.91 | 26.20 | 3BS | kdk97 | 61 |
| Orion dwarf | 05:45:02.1 | +05:04:09 | −12 | 0.60 | +29 | 0.84 | 28.66 | 3BS | km96 | 177 |
| DDO 47 | 07:41:55.4 | +16:48:09 | +19 | 0.00 | 0 | 0.04 | 28.39 | TRGB | kms03 | 79 |
| UGC 4115 | 07:57:02.3 | +14:23:23 | +21 | 0.60 | −39 | 0.03 | 28.53 | TRGB | kms03 | 112 |
| DDO 53 | 08:34:07.2 | +66:10:54 | +35 | (b) | (b) | 0.04 | 27.58 | TRGB | kdg02 | (b) |
| UGC 4483 | 08:37:03.4 | +69:46:35 | +34 | 0.50 | −10 | 0.04 | 27.37 | TRGB | dmk01 | 56 |
| UGC 4998 | 09:25:12.4 | +68:23:00 | +39 | 0.30 | +75 | 0.07 | 29.95 | SBF | jrt01 | – |
| Holmberg I | 09:40:32.3 | +71:10:56 | +39 | (a) | (a) | 0.06 | 27.74 | TRGB | kdg02 | (a) |
| Leo A | 09:59:26.4 | +30:44:47 | +52 | (a) | (a) | 0.02 | 24.35 | TRGB | tgc98 | (a) |
| Sextans B | 10:00:00.1 | +05:19:56 | +44 | (a) | (a) | 0.04 | 25.53 | TRGB | ksm02 | (a) |
| UGC 5423 | 10:05:31.0 | +70:21:53 | +41 | 0.50 | −40 | 0.09 | 28.11 | 3BS | skt99 | 67 |
| Sextans A | 10:11:00.8 | −04:41:34 | +40 | (a) | (a) | 0.05 | 25.55 | TRGB | dss03 | (a) |
| UGC 5692 | 10:30:34.8 | +70:37:17 | +42 | 0.30 | −19 | 0.05 | 27.83 | TRGB | kdg02 | – |
| UGC 5848 | 10:44:23.0 | +56:25:17 | +53 | 0.40 | −65 | 0.01 | 29.70 | HUB | stm92 | 145 |
| UGC 5979 | 10:52:41.1 | +67:59:20 | +45 | 0.20 | +80 | 0.02 | 30.55 | HUB | hs98 | 104 |
| UGC 6456 | 11:27:59.4 | +78:59:37 | +37 | 0.30 | 0 | 0.04 | 28.03 | TRGB | mdm02 | 58 |
| Markarian 178 | 11:33:28.8 | +49:14:22 | +63 | 0.50 | −49 | 0.02 | 27.78 | TRGB | ksd03 | 47 |
| NGC 3738 | 11:35:48.8 | +54:31:28 | +59 | 0.30 | −24 | 0.01 | 28.27 | TRGB | ksd03 | 109 |
| NGC 3741 | 11:36:06.1 | +45:17:13 | +66 | 0.26 | +23 | 0.03 | 27.24 | TRGB | ksd03 | – |
| NGC 4163 | 12:12:09.1 | +36:10:07 | +78 | 0.30 | +11 | 0.02 | 27.30 | 3BS | tk98 | 47 |
| NGC 4190 | 12:13:44.4 | +36:38:05 | +78 | 0.35 | +21 | 0.03 | 27.20 | 3BS | tk98 | 67 |
| Markarian 209 | 12:26:17.0 | +48:29:39 | +68 | 0.00 | +90 | 0.02 | 28.37 | 3BS | mkg97 | 76 |
| PGC 42275 | 12:38:40.0 | +32:46:01 | +84 | (b) | (b) | 0.02 | 27.06 | 3BS | mkt98 | (b) |
| NGC 4789A | 12:54:05.4 | +27:08:57 | +89 | 0.60 | +26 | 0.01 | 27.80 | 3BS | mkt98 | 98 |
| GR 8 | 12:58:39.8 | +14:13:05 | +77 | 0.20 | +61 | 0.03 | 26.52 | TRGB | dsg98 | 44 |
| DDO 167 | 13:13:22.7 | +46:19:13 | +70 | 0.40 | −30 | 0.01 | 27.93 | TRGB | ksd03 | 44 |
| UGC 8508 | 13:30:44.4 | +54:54:41 | +61 | 0.45 | −60 | 0.02 | 26.87 | TRGB | ksm02 | 63 |
| NGC 5264 | 13:41:37.0 | −29:54:41 | +32 | 0.30 | +56 | 0.06 | 28.11 | TRGB | ksd02 | 52 |
| Holmberg IV | 13:54:46.0 | +53:54:20 | +61 | 0.70 | +21 | 0.02 | 29.22 | 3BS | mak99 | 99 |
| DDO 187 | 14:15:56.5 | +23:03:20 | +70 | 0.40 | +46 | 0.03 | 26.78 | TRGB | atk00 | 49 |

[a] Galaxy larger than FOV

[b] No isophotes detected

Note. — Col. (1): Galaxy name; Col. (2): Right ascension (ellipse fit of the unresolved component in $K_s$); Col. (3): Declination (idem); Col. (4): Galactic latitude (degrees, from NED); Col. (5): Ellipticity (1-b/a); Col. (6): Position angle (from North to East); Col. (7): Optical depth at 1 micron calculated by YES; Col. (8): Distance modulus; Col. (9): Method of determining distance. MEM: Group Membership; TRGB: Tip of the Red Giant Branch; 3BS: Brightest Blue Stars; SBF: Surface Brightness Fluctuations; HUB: Hubble law; Col. (10): Reference to distance determinations; Col. (11): Width of the HI line profile measured in km/s at 20% of the peak;

References. — dsg98: Dohm-Palmer et al. 1998, dss03: Dolphin et al. 2003, dmk01: Dolphin et al. 2001, fin02: Fingerhut 2002, hs98: Huchtmeier and Skillman 1998, jrt01: Jerjen et al. 2001, kma99: Karachentsev, Makarova and Andersen 1999, kms03: Karachentsev et al. 2003a, ksd03: Karachentsev et al. 2003b, ksm02: Karachentsev et al. 2002a, ksd02: Karachentsev et al. 2002b, kdg02: Karachentsev et al. 2002c, kdk97: Karachentsev et al. 1997, km96: Karachentsev and Musella 1996, mak99: Makarova 1999, mk03: Makarova and Karachentsev 2003, mkt98: Makarova et al. 1998, mkg97: Makarova, Karachentsev and Georgiev 1997, mdm02: Méndez et al. 2002, stm92: Schneider et al. 1992, skt99: Sharina, Karachentsev and Tikhonov 1999, tk98: Tikhonov and Karachentsev 1998, tgc98: Tolstoy et al. 1998.

Table 2. Observing Log

| Galaxy | Observatory | Date (UT) | Filter | Exp Time (sec) |
|---|---|---|---|---|
| Cassiopeia 1 | OAN-SPM | 2002 Mar 1 | $K_s$ | 1800 |
|  | OAN-SPM | 2003 Mar 15 | $J$ | 1200 |
| MB 1 | CFHT | 2002 Mar 2 | $K_s$ | 600 |
|  | CFHT | 2002 Mar 2 | $J$ | 300 |
| Cam A | CFHT | 2002 Mar 3 | $K_s$ | 600 |
|  | CFHT | 2002 Mar 3 | $J$ | 300 |
| NGC 1569 | CFHT | 2002 Mar 1 | $K_s$ | 600 |
|  | CFHT | 2002 Mar 1 | $J$ | 300 |
|  | OAN-SPM | 2002 Feb 28 | $K_s$ | 2160 |
|  | OAN-SPM | 2002 Feb 28 | $J$ | 1020 |
| UGCA 92 | CFHT | 2002 Mar 2 | $K_s$ | 600 |
|  | CFHT | 2002 Mar 2 | $J$ | 300 |
| Orion Dwarf | CFHT | 2002 Mar 3 | $K_s$ | 600 |
|  | CFHT | 2002 Mar 3 | $J$ | 300 |
| DDO 47 | CFHT | 2002 Mar 2 | $K_s$ | 600 |
|  | CFHT | 2002 Mar 2 | $J$ | 300 |
| UGC 4115 | CFHT | 2002 Mar 2 | $K_s$ | 600 |
|  | CFHT | 2002 Mar 2 | $J$ | 300 |
| DDO 53 | CFHT | 2002 Mar 1 | $K_s$ | 600 |
|  | CFHT | 2002 Mar 1 | $J$ | 300 |
|  | OAN-SPM | 2001 Mar 2 | $K_s$ | 2760 |
| UGC 4483 | CFHT | 2004 Feb 24 | $K_s$ | 525 |
|  | CFHT | 2004 Feb 24 | $J$ | 300 |
|  | OAN-SPM | 2004 Feb 12,17 | $K_s$ | 4320 |
|  | OAN-SPM | 2004 Feb 12,16 | $J$ | 2400 |
| UGC 4998 | OAN-SPM | 2004 Feb 13 | $K_s$ | 2340 |
|  | OAN-SPM | 2004 Feb 13 | $J$ | 1200 |
| Holmberg I | CFHT | 2002 Mar 3 | $K_s$ | 600 |
|  | CFHT | 2002 Mar 3 | $J$ | 300 |
| Leo A | CFHT | 2002 Mar 3 | $K_s$ | 600 |
|  | CFHT | 2002 Mar 3 | $J$ | 300 |
| Sextans B | CFHT | 2002 Mar 3 | $K_s$ | 600 |
|  | CFHT | 2002 Mar 3 | $J$ | 300 |
| UGC 5423 | CFHT | 2004 Mar 7 | $K_s$ | 525 |
|  | CFHT | 2004 Mar 7 | $J$ | 300 |
| Sextans A | CFHT | 2004 Mar 8 | $K_s$ | 600 |
|  | CFHT | 2004 Mar 8 | $J$ | 300 |
| UGC 5692 | CFHT | 2002 Mar 3 | $K_s$ | 600 |
|  | CFHT | 2002 Mar 3 | $J$ | 300 |
| UGC 5848 | OAN-SPM | 2003 Mar 17,19 | $K_s$ | 2040 |
|  | OAN-SPM | 2003 Mar 15 | $J$ | 1200 |
| UGC 5979 | OAN-SPM | 2004 Feb 15 | $K_s$ | 1140 |
|  | OAN-SPM | 2004 Feb 15 | $J$ | 1200 |
| UGC 6456 | CFHT | 2002 Mar 1 | $K_s$ | 600 |
|  | CFHT | 2002 Mar 1 | $J$ | 300 |
| Markarian 178 | CFHT | 2002 Mar 2 | $K_s$ | 600 |
|  | CFHT | 2002 Mar 2 | $J$ | 300 |
| NGC 3738 | CFHT | 2002 Mar 1 | $K_s$ | 600 |

Table 2—Continued

| Galaxy | Observatory | Date (UT) | Filter | Exp Time (sec) |
|---|---|---|---|---|
|  | CFHT | 2002 Mar 1 | $J$ | 300 |
| NGC 3741 | CFHT | 2002 Mar 3 | $K_s$ | 600 |
|  | CFHT | 2002 Mar 3 | $J$ | 300 |
| NGC 4163 | CFHT | 2002 Mar 3 | $K_s$ | 600 |
|  | CFHT | 2002 Mar 3 | $J$ | 300 |
| NGC 4190 | CFHT | 2002 Mar 3 | $K_s$ | 600 |
|  | CFHT | 2002 Mar 3 | $J$ | 300 |
| Markarian 209 | CFHT | 2002 Mar 1 | $K_s$ | 600 |
|  | CFHT | 2002 Mar 1 | $J$ | 300 |
| PGC 42275 | CFHT | 2002 Mar 2 | $K_s$ | 600 |
|  | CFHT | 2002 Mar 2 | $J$ | 300 |
| NGC 4789A | CFHT | 2002 Mar 1 | $K_s$ | 600 |
|  | CFHT | 2002 Mar 1 | $J$ | 300 |
| GR 8 | CFHT | 2002 Mar 2, 3 | $K_s$ | 1200 |
|  | CFHT | 2002 Mar 2, 3 | $J$ | 600 |
| DDO 167 | CFHT | 2002 Mar 1 | $K_s$ | 600 |
|  | CFHT | 2002 Mar 1 | $J$ | 300 |
| UGC 8508 | OAN-SPM | 2002 Mar 1 | $K_s$ | 2520 |
|  | OAN-SPM | 2002 Mar 1 | $J$ | 900 |
| NGC 5264 | CFHT | 2002 Mar 2 | $K_s$ | 600 |
|  | CFHT | 2002 Mar 2 | $J$ | 300 |
| Holmberg IV | CFHT | 2002 Mar 3 | $K_s$ | 600 |
|  | CFHT | 2002 Mar 3 | $J$ | 300 |
| DDO 187 | CFHT | 2002 Mar 1, 2 | $K_s$ | 1200 |
|  | CFHT | 2002 Mar 1, 2 | $J$ | 600 |
|  | OAN-SPM | 2001 Mar 3 | $K_s$ | 2940 |

Table 3. Photometric Parameters[d]

| Galaxy | Filter | $Res/Tot$ (%) | $m_I$ (mag) | $m_T$ (mag) | $m_S$ (mag) | $\mu_0$ (mag/$''^2$) | $r_0$ ($''$) | $r_{22(23)}$ ($''$) |
|---|---|---|---|---|---|---|---|---|
| (1) | (2) | (3) | (4) | (5) | (6) | (7) | (8) | (9) |
| Casiopeia 1 | $J$ | (c) | 11.04 | (c) | 10.81 | 20.49 | 25.4 | 76.2 |
| | $K_s$ | (c) | 9.97 | (c) | 9.75 | 19.51 | 26.3 | 78.7 |
| MB 1 | $J$ | 8.4 | 11.79 | 11.68 | 11.67 | 20.10 | 22.7 | 76.2 |
| | $K_s$ | 3.1 | 10.52 | 10.46 | 10.40 | 18.98 | 24.3 | 84.3 |
| Cam A | $J$ | (a) | (a) | (a) | (a) | (a) | (a) | (a) |
| | $K_s$ | (a) | (a) | (a) | (a) | (a) | (a) | (a) |
| NGC 1569 | $J$ | 3.4 | 9.06 | 8.94 | 9.37 | 17.99 | 21.0 | 111.7 |
| | $K_s$ | 2.7 | 8.12 | 8.04 | 8.45 | 17.09 | 21.2 | 110.8 |
| UGCA 92 | $J$ | 9.9 | 12.58 | 12.38 | 12.27 | 22.05 | 37.6 | 57.3 |
| | $K_s$ | 5.0 | 11.21 | 11.12 | 10.82 | 20.83 | 41.7 | 72.8 |
| Orion dwarf | $J$ | 4.4 | 11.75 | 11.70 | 11.72 | 20.48 | 26.3 | 79.4 |
| | $K_s$ | 5.6 | 10.96 | 10.90 | 10.98 | 19.56 | 24.3 | 71.4 |
| DDO 47 | $J$ | 4.1 | 13.57 | 13.50 | 13.38 | 22.07 | 16.1 | 24.3 |
| | $K_s$ | 4.6 | 13.46 | 13.40 | 13.36 | 21.79 | 14.3 | 9.1 |
| UGC 4115 | $J$ | 8.3 | 13.22 | 13.12 | 13.10 | 21.11 | 18.6 | 45.2 |
| | $K_s$ | 4.3 | 12.18 | 12.13 | 11.94 | 20.25 | 21.4 | 49.1 |
| DDO 53 | $J$ | (b) | (b) | (b) | (b) | (b) | (b) | (b) |
| | $K_s$ | (b) | (b) | (b) | (b) | (b) | (b) | (b) |
| UGC 4483 | $J$ | 4.1 | 14.27 | 14.14 | 13.81 | 21.98 | 18.0 | 28.6 |
| | $K_s$ | 1.7 | 13.36 | 13.33 | 12.71 | 20.62 | 15.9 | 31.0 |
| UGC 4998 | $J$ | (c) | 11.85 | (c) | 12.01 | 20.12 | 13.5 | 45.3 |
| | $K_s$ | (c) | 11.11 | (c) | 11.44 | 19.43 | 13.2 | 40.3 |
| Holmberg I | $J$ | (a) | (a) | (a) | (a) | (a) | (a) | (a) |
| | $K_s$ | (a) | (a) | (a) | (a) | (a) | (a) | (a) |
| Leo A | $J$ | (a) | (a) | (a) | (a) | (a) | (a) | (a) |
| | $K_s$ | (a) | (a) | (a) | (a) | (a) | (a) | (a) |
| Sextans B | $J$ | (a) | (a) | (a) | (a) | (a) | (a) | (a) |
| | $K_s$ | (a) | (a) | (a) | (a) | (a) | (a) | (a) |
| UGC 5423 | $J$ | 1.2 | 13.20 | 13.18 | 13.07 | 20.67 | 13.8 | 39.1 |
| | $K_s$ | 0.5 | 12.10 | 12.09 | 11.97 | 19.75 | 15.0 | 41.5 |
| Sextans A | $J$ | (a) | (a) | (a) | (a) | (a) | (a) | (a) |
| | $K_s$ | (a) | (a) | (a) | (a) | (a) | (a) | (a) |
| UGC 5692 | $J$ | 2.0 | 10.98 | 10.96 | 10.87 | 20.43 | 28.8 | 88.1 |
| | $K_s$ | 1.1 | 10.31 | 10.29 | 10.22 | 19.56 | 26.0 | 76.4 |
| UGC 5848 | $J$ | (c) | 12.34 | (c) | 12.35 | 20.84 | 19.0 | 50.7 |
| | $K_s$ | (c) | 11.58 | (c) | 11.59 | 19.90 | 17.5 | 45.8 |
| UGC 5979 | $J$ | (c) | 12.57 | (c) | 13.00 | 21.44 | 16.1 | 34.0 |
| | $K_s$ | (c) | 11.66 | (c) | 12.06 | 20.46 | 15.8 | 33.0 |
| UGC 6456 | $J$ | 4.5 | 13.28 | 13.23 | 13.44 | 20.89 | 10.9 | 28.7 |
| | $K_s$ | 3.2 | 12.52 | 12.48 | 12.54 | 20.17 | 11.8 | 28.0 |
| Markarian 178 | $J$ | 3.2 | 12.46 | 12.42 | 12.35 | 19.72 | 12.4 | 46.1 |
| | $K_s$ | 2.3 | 11.88 | 11.85 | 11.74 | 19.15 | 12.6 | 41.9 |
| NGC 3738 | $J$ | 5.8 | 10.46 | 10.37 | 10.54 | 19.26 | 19.5 | 80.7 |
| | $K_s$ | 3.7 | 9.61 | 9.54 | 9.68 | 18.41 | 19.7 | 78.7 |
| NGC 3741 | $J$ | 2.9 | 12.90 | 12.84 | 12.81 | 20.63 | 12.6 | 36.1 |
| | $K_s$ | 2.7 | 12.32 | 12.26 | 12.22 | 20.04 | 12.5 | 31.3 |
| NGC 4163 | $J$ | 3.0 | 11.63 | 11.57 | 11.66 | 19.88 | 15.6 | 55.5 |

Table 3—Continued

| Galaxy | Filter | $Res/Tot$ (%) | $m_I$ (mag) | $m_T$ (mag) | $m_S$ (mag) | $\mu_0$ (mag/$''^2$) | $r_0$ ($''$) | $r_{22(23)}$ ($''$) |
|---|---|---|---|---|---|---|---|---|
| (1) | (2) | (3) | (4) | (5) | (6) | (7) | (8) | (9) |
|  | $K_s$ | 5.8 | 10.93 | 10.88 | 10.96 | 19.30 | 16.4 | 52.1 |
| NGC 4190 | $J$ | 2.7 | 11.42 | 11.38 | 11.32 | 19.78 | 18.0 | 65.8 |
|  | $K_s$ | 2.9 | 10.84 | 10.79 | 10.74 | 19.12 | 17.3 | 58.0 |
| Markarian 209 | $J$ | 10.0 | 13.71 | 13.60 | 13.77 | 20.69 | 7.1 | 20.1 |
|  | $K_s$ | 5.7 | 12.56 | 12.51 | 12.70 | 20.03 | 8.6 | 21.6 |
| PGC 42275 | $J$ | (b) | (b) | (b) | (b) | (b) | (b) | (b) |
|  | $K_s$ | (b) | (b) | (b) | (b) | (b) | (b) | (b) |
| NGC 4789A | $J$ | 9.3 | 13.40 | 13.26 | 13.15 | 21.91 | 26.4 | 43.7 |
|  | $K_s$ | 4.8 | 12.41 | 12.33 | 12.17 | 21.13 | 28.9 | 41.5 |
| GR 8 | $J$ | 5.5 | 13.38 | 13.29 | 13.23 | 22.00 | 18.7 | 29.5 |
|  | $K_s$ | 5.7 | 13.04 | 12.90 | 12.92 | 21.37 | 16.2 | 19.1 |
| DDO 167 | $J$ | 2.1 | 13.93 | 13.91 | 14.05 | 22.29 | 17.0 | 21.5 |
|  | $K_s$ | 1.4 | 13.47 | 13.45 | 13.37 | 21.65 | 17.3 | 14.6 |
| UGC 8508 | $J$ | (c) | 11.97 | (c) | 12.02 | 20.47 | 19.6 | 59.1 |
|  | $K_s$ | (c) | 11.57 | (c) | 11.44 | 19.96 | 19.4 | 49.9 |
| NGC 5264 | $J$ | 2.5 | 10.31 | 10.28 | 10.26 | 19.50 | 24.8 | 97.3 |
|  | $K_s$ | 1.5 | 9.52 | 9.50 | 9.48 | 18.80 | 25.7 | 93.6 |
| Holmberg IV | $J$ | 4.8 | 12.72 | 12.66 | 12.43 | 21.12 | 29.5 | 71.3 |
|  | $K_s$ | 2.9 | 11.07 | 11.04 | 11.03 | 20.25 | 37.8 | 86.5 |
| DDO 187 | $J$ | 4.3 | 13.08 | 12.98 | 12.93 | 22.10 | 26.0 | 37.7 |
|  | $K_s$ | 9.3 | 12.60 | 12.46 | 12.45 | 21.72 | 27.2 | 5.8 |

[a] Galaxy larger than FOV

[b] No unresolved component detected

[c] Observed at OAN-SPM; no stars resolved

[d] No corrections for Galactic extinction applied

Note. — Col. (1): Galaxy name; Col. (2): Filter; Col. (3): Resolved/Total Flux; Col. (4): Isophotal magnitude of the unresolved component; Col. (5): Total magnitude (unresolved plus resolved); Col. (6): Sech magnitude (from sech law); Col. (7): Central surface brightness (of sech law); Col. (8): Scale length (of sech law); Col. (9): Semimajor axis $r_{22}$ corresponding to $K_s = 22$ mag arcsec$^{-2}$; semimajor axis $r_{23}$ corresponding to $J = 23$ mag arcsec$^{-2}$.

Table 4.  Comparison with 2MASS

| Galaxy | $\Delta\alpha$ (arcsec) | $\Delta\delta$ (arcsec) | $\Delta$e | $\Delta$PA (deg) | $\Delta m_J$ (mag) | $\Delta m_K$ (mag) |
| --- | --- | --- | --- | --- | --- | --- |
| (1) | (2) | (3) | (4) | (5) | (6) | (7) |
| NGC 1569  | −2.2 | −2.4  | +0.09 | +1.8  | −0.11 | −0.16 |
| NGC 5264  | −5.0 | −6.3  | +0.00 | +14.0 | +1.06 | +1.32 |
| NGC 4190  | −4.2 | −12.1 | +0.41 | +9.0  | +1.67 | +1.75 |
| NGC 4789A | −2.5 | +1.5  | −0.60 | +63.8 | +2.98 | +1.53 |

Note. — Col. (1): Galaxy name; Col. (2): Difference in right ascension (2MASS minus ours); Col. (3): Declination (idem); Col. (4): Ellipticity (idem); Col. (5): Position angle (idem); Col. (6): Total magnitude J (idem); Col. (7): Total magnitude $K_s$ (idem);

Full version including figures available at

http://aries.phys.yorku.ca/~ovidiuv/Paper2.ps.gz

(10 MB gzip - sorry, too big for arXiv.org to store)

Fig. 1.— Field dwarf galaxies observed at CFHT for which an unresolved component was detected. Left panel: $K_s$ images (north is up, east to left, $3\rlap{.}'6 \times 3\rlap{.}'6$, binned 4x4). Right panel: surface brightness profiles in $J$ and $K_s$ (upper graphs), and $J - K_s$ color profiles (lower graphs).

Fig. 1.— B

Fig. 1.— C

Fig. 1.— D

Fig. 1.— E

Fig. 1.— F

Fig. 1.— G

Fig. 1.— H

Fig. 2.— $K_s$ images of field dwarf galaxies observed at CFHT larger than the field of view or for which an unresolved component was not detected. North is up, east to left, $3'\!.6 \times 3'\!.6$, binned 4x4.

Fig. 3.— Field dwarf galaxies observed at OAN-SPM. Left panel: $K_s$ images (north is up, east to left $3'\!.6 \times 3'\!.6$). Right panel: surface brightness profiles in $J$ and $K_s$ (upper graphs), and $J - K_s$ color profiles (lower graphs).

Fig. 3.— B

Fig. 3.— C

Fig. 4.— (a) $K_s$ image of the dwarf irregular galaxy NGC 1569 observed with CFHT-IR. (b) The unresolved part of NGC 1569, namely what is left after removing sources with KILLALL. North is up, east to left, FOV $3'\!.6 \times 3'\!.6$.

Fig. 5.— Residual magnitudes (recovered - injected) of the recovered stars from the 600 artificial stars added to the image of NGC 1569, plotted as a function of the injected magnitudes.

Fig. 6.— Color magnitude diagrams of the resolved stars in the galaxies observed at CFHT. The horizontal solid lines are placed at the apparent $K_s$ magnitude corresponding to $M_{K_s} = -7.5$, taking into account the distance and extinction in Table 1. The first number at the bottom right of each CMD is the number of stars plotted. The second number is the percentage of stars expected to be foreground contaminants. Percentages are not given for galaxies larger than the field of view.

Fig. 6.— B

Fig. 6.— C

Fig. 6.— D

Fig. 7.— Color magnitude diagrams for the stars outside the selected star catalogs (i.e., part of the Milky Way). The first number at the bottom right of each CMD is the number of stars plotted. The second number gives the percentage of the stars expected to be in the foreground (Galactic contaminants), based on the galaxy's area relative to the field area from which the stars plotted here were extracted.

Fig. 8.— Uncertainties in magnitudes for the stars in the selected star catalog for NGC 1569, as a function of apparent magnitude. An average error of about 0.2 mag can be observed in $J$ and about 0.25 mag in $K_s$, with the maximum reaching about 0.5 mag at the faint end in both bands.

Fig. 9.— The residuals between the isophotal magnitudes $m_I$ and the sech magnitudes $m_S$ showing the goodness of the "sech" fit to the surface brighness profiles in $K_s$. The three UGC galaxies are faint, while NGC 1569 has an anomalously bright core due to its starburst. All other unresolved components could be fitted within 0.2 mag in total magnitude.

Fig. 10.— The fraction of the light of a burst of star formation contributed by stars brighter than $M_{K_s} = -7.5$ with respect to the total flux from all stars, as a function of the age. The plot is derived from isochrones generated using a Salpeter initial mass function and an isochrone with metallicity Z=0.0004.

Fig. 11.— The correlation between the scale length and the absolute sech magnitude in $K_s$. The three labeled galaxies at top-left are expected to have larger errors, while NGC 1569 is known to be a starburst. The dotted line represents a least square fit to the rest of the points, whose coefficients are given in Eq. 15. The thin curve shows a locus of constant central surface brightness, specifically $\mu_{0K} = -19$ mag arcsec$^{-2}$.

Fig. 12.— The correlation between the semimajor radius $r_{22}$ and the absolute sech magnitude in $K_s$. The three labeled galaxies at top-left are expected to have larger errors, while NGC 1569 is known to host a starburst. The dotted line represents a least square fit to the rest of the points, whose coefficients are given in Eq. (16).

Fig. 13.— The correlation between the central surface brightness $\mu_0$ and the absolute sech magnitude in $K_s$. The three labeled galaxies at left are expected to have larger errors, while NGC 1569 is known to have a central starburst. The thin line shows a locus of constant scale length, specifically $r_{0K} = 0.4$ kpc.

Fig. 14.— The correlation between the total color and the absolute total magnitude in $K_s$. UGC 5979 and Ho IV have approximative distances, also Ho IV is larger than the field of view, so its sech magnitude might have a larger error. NGC 5264 might be a dwarf spiral, in which case tilt correction need to be applied to its $W_{20}$.

Fig. 15.— $M_K$ as a function of the logarithm of the neutral hydrogen line width, $W_{20}$ (km/s). Sech magnitudes are plotted for our field dIs (filled circles), while total magnitudes are plotted for the Virgo dwarfs examined by Pierini & Tuffs (1999) (open circles).

Fig. 16.— The sech magnitude in $K_s$ versus the logarithm of the neutral hydrogen line width, $W_{20}$ (km/s) and the central surface brightness, $\mu_0$. Note the reduced scatter compared with the Tully-Fisher relation shown in the previous figure. The error bar at the upper left of the diagram shows the typical error in $M_K$ arising from the uncertainty in TRGB distance. The photometry for Ho IV and the distances for UGC 5979 and NGC 5264 are questionable.